\documentclass{revtex4}
\usepackage{graphicx}
\usepackage{mathbbol}
\def\be{\begin{equation}}
\def\ee{\end{equation}}
\def\bea{\begin{eqnarray}}
\def\eea{\end{eqnarray}}
\def\l{\left}
\def\r{\right}
\def\fraction{\displaystyle\frac}

\begin{document}

\title{Introduction to spectral methods} 
\author{Philippe Grandcl\'ement}
\affiliation{Laboratoire Univers et ses Th\'eories, 
Observatoire de Paris,
5 place J. Janssen, 
92195 Meudon Cedex, France}
\begin{abstract}
This proceeding is intended to be a first introduction to spectral methods. It is written around some simple problems that are solved explicitly and in details and that aim at demonstrating the power of those methods. The mathematical foundation of the spectral approximation is first introduced, based on the Gauss quadratures. The two usual basis of Legendre and Chebyshev polynomials are then presented. The next section is devoted to one dimensional equation solvers using only one domain. Three different methods are described. Techniques using several domains are shown in the last section of this paper and their various merits discussed.
\end{abstract}
\maketitle
\section{Introduction}
This proceeding constitutes a short introduction to {\em spectral methods}. The aim is not to present an exhaustive mathematical presentation of those methods. Numerous books can be consulted for this purpose (see the bibliography for a sample of them). The following material should instead be considered as a toolkit for implementing simple spectral methods solvers. Thus, a particular emphasize will be put on practical examples. For the sake of simplicity, we will restrict ourselves to one dimensional solvers.

Spectral methods are just one of the many ways to represent a function on a computer. The basic idea of all numerical techniques is to approximate any function by polynomials, those being the only functions than a computer can exactly calculate. So a function $u$ will be approximate by $\hat{u} = \sum_{n=0}^N \hat{u}_n \Phi_n$ where the $\Phi_n$ are polynomials and called the {\em trial functions}. Depending on the choice of trial functions, one can generate various classes of numerical techniques. For example, the {\em finite difference} schemes are obtained by choosing local polynomials of low degree. In all the following, we will be interested in spectral methods where the 
$\Phi_n$ are global polynomials, typically Legendre of Chebyshev ones. If spectral methods are basically more evolved than finite difference schemes, they have long prove their ability to tackle a wide variety of problems. In particular, they allow to reach very good accuracy with only moderate computational resources. 

In the first part of this article, we present the basic properties of spectral methods. After a brief overview of the foundations of the spectral expansion, we show the power of the method on simple but non-trivial examples. In particular, we show that, for ${{\mathcal C}^\infty}$ functions, the error decays exponentially, as one increases the degree of the approximation. We then present some basic features of the two type of polynomials that are used in all the rest of this article, Legendre and Chebyshev polynomials.

The second section is devoted to the actual implementation of partial differential equation solvers. We restrict ourselves to one dimensional equations on a bounded domain (i.e. $x \in \l[-1, 1\r]$). Three different types of solvers are presented:~the {\em Tau-method}, the {\em collocation method} and the {\em Galerkin method}. They are explicitly implemented on a simple example.

The last part of this work is concerned with the implementation of {\em multi-domain} solvers. After having explained why this is often a valuable tool for the physicist, we present, once again, three different methods, all of them being tested an implemented on a simple, one dimensional equation, with a decomposition of space using two domains.

\section{Foundations of spectral methods}
\subsection{Spectral expansion}
\subsubsection{Orthogonal projection}
Let us consider an interval $\Lambda = \l[x_{\rm min}, x_{\rm max}\r]$. In order to talk about basis, one needs to define a scalar product on $\Lambda$. If $w$ is a positive function on $\Lambda$, one can define the scalar product of two functions $f$ and $g$, with respect to the {\em measure} $w$ as being 
\be
\label{scal_prod}
\l(f,g\r)_w = \int_\Lambda f\l(x\r) g\l(x\r) w\l(x\r) {\rm d} x.
\ee
Using this scalar product, one can find a set of orthogonal polynomials $p_n$, each of them of degree $n$. The set composed of those polynomials, up to a given degree $N$ is a basis of $\mathbb{P}_N$.

\begin{figure}
\centerline{
\includegraphics[height=5.5cm]{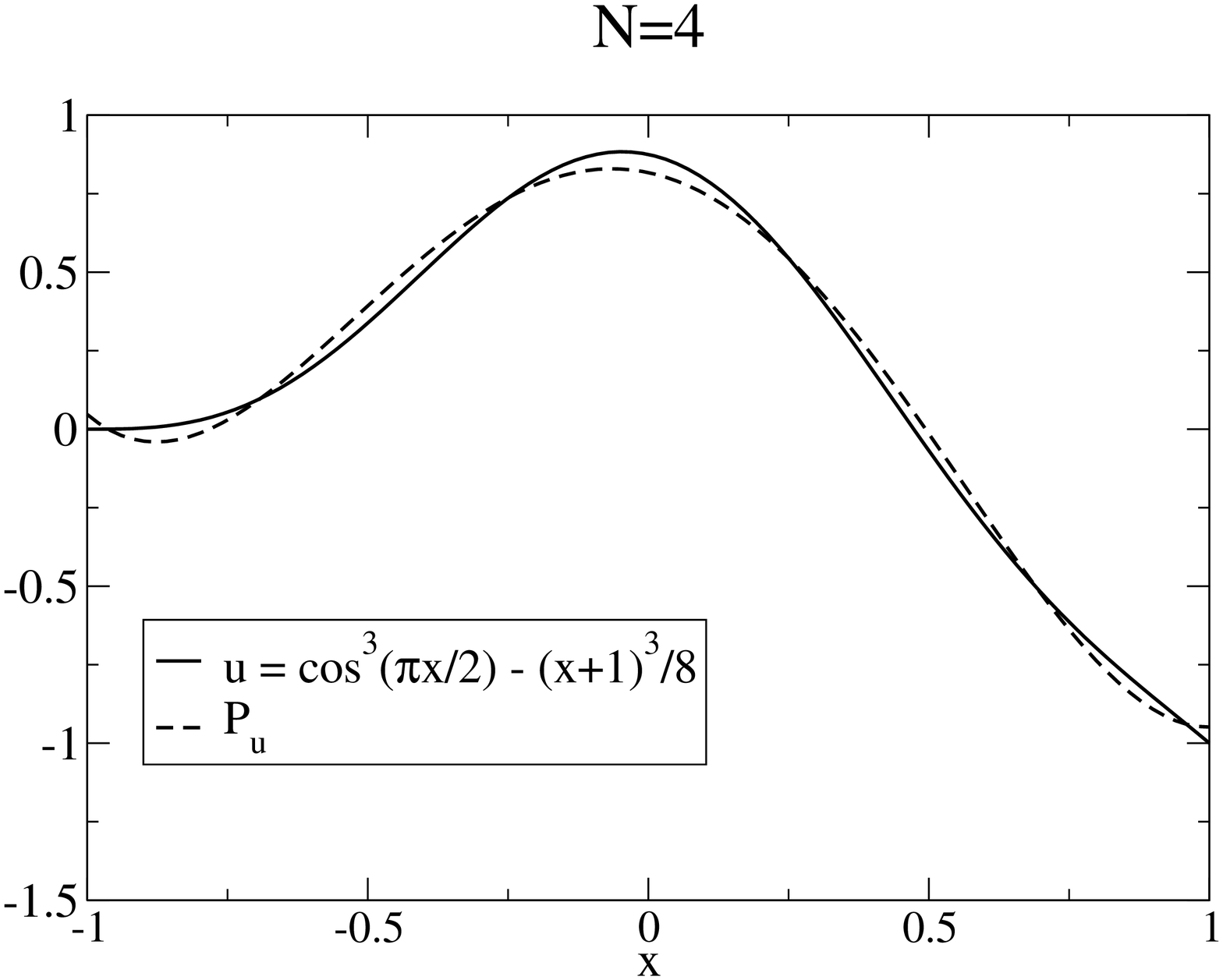}
\includegraphics[height=5.5cm]{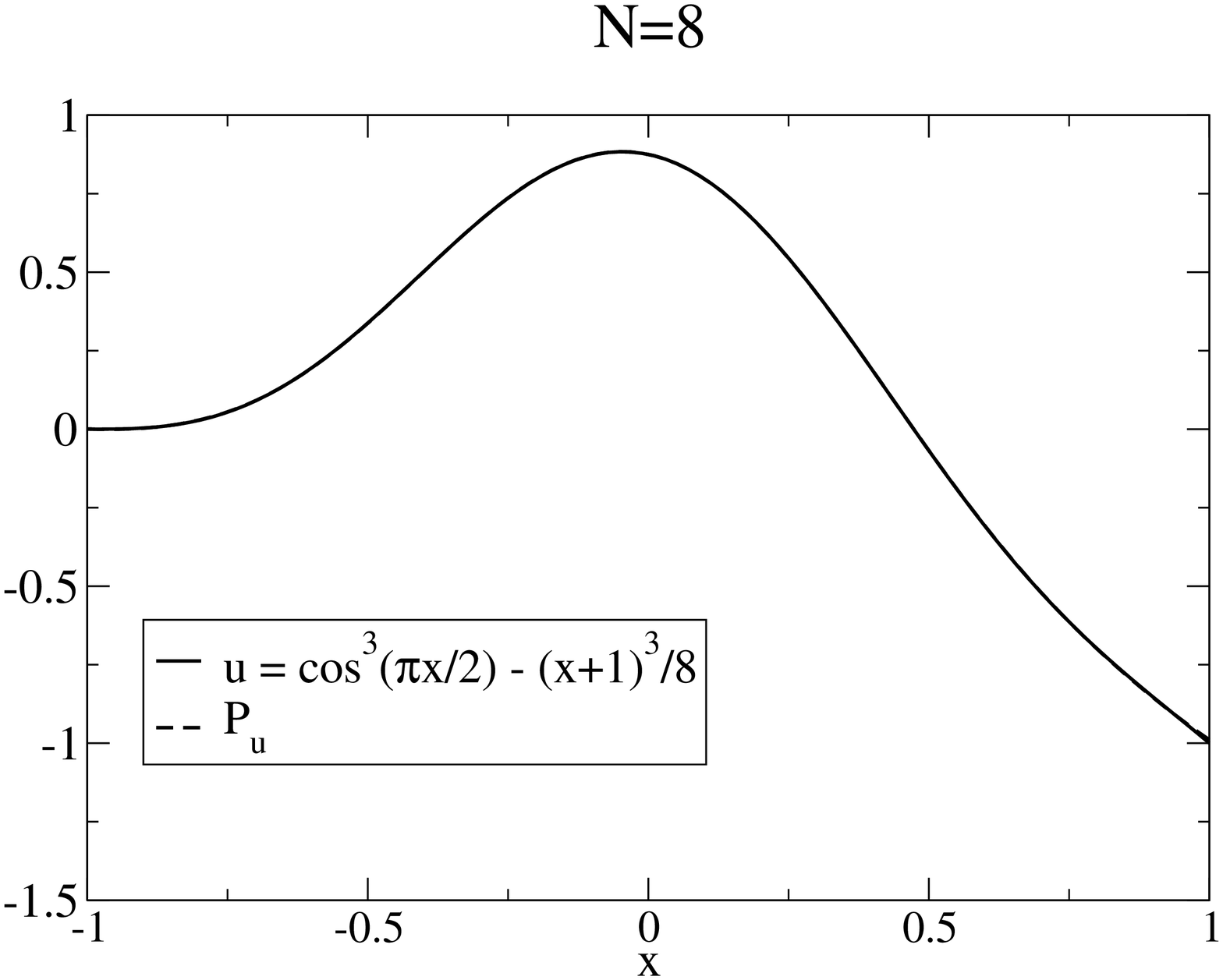}}
\caption{\label{f:proj}
$u= \cos^3\l(\pi x/2\r)-\l(x+1\r)^3/8$ and its projection of Chebyshev polynomials of degree smaller than $4$ (left panel) and $8$ (right panel).
}
\end{figure}

One can then hope to represent any function $u$ on $\Lambda$ by its projection on the polynomials $p_n$. Doing so, we define the projection of $u$ simply by 
\be
\label{proj}
P_Nu = \sum_{n=0}^N \hat{u}_n p_n\l(x\r),
\ee
where the coefficients of the projections are given by $\hat{u}_n = \fraction{\l(u, p_n\r)}{\l(p_n, p_n\r)}$. The difference between $u$ and its projection is called the {\em truncation error} and one can show that it goes to zero when the order of the approximation increases : 
\be
\l\|u-P_nu\r\| \longrightarrow 0 \quad {\rm when} \quad N \longrightarrow \infty.
\ee
This convergence is illustrated on Fig. \ref{f:proj}, where a test function $u$ is plotted, with its projection for $N=4$ and $N=8$. Chebyshev polynomials are used. Let us note that, even if $u$ is not a polynomial function, no discrepancy can be seen by eye, for $N$ as small as 8.

This seems very appealing but for the fact that one needs to calculate the $\hat{u}_n$ by computing integrals like 
$\displaystyle\int_\Lambda u\l(x\r) p_n\l(x\r) w\l(x\r) {\rm d} x$. If one needs to evaluate $u$ at a lot of points in order to do so (like when using standard Newton method), one would lose all the advantage of using spectral expansion.

\subsubsection{Gauss quadratures}
The solution to this problem is given by the Gauss quadratures. The theorem can be stated as follows. 

{\em There exist $N+1$ positive reals $w_n$ and $N+1$ reals $x_n$ in $\Lambda$ such that :}
\be
\label{gauss}
\forall f\in {\mathbb P}_{2N+\delta}, \quad \int_\Lambda f\l(x\r) w\l(x\r) {\rm d}x = 
\sum_{n=0}^N f\l(x_n\r) w_n.
\ee
The $w_n$ are called the {\em weights} and the $x_n$ the {collocation points}. 
The exact degree of applicability depends on the quadrature. The three usual choices are :
\begin{itemize}
\item Gauss : $\delta =1$
\item Gauss-Radau : $\delta =0$ and $x_0 = x_{\rm min}$.
\item Gauss-Lobatto : $\delta = -1$ and $x_0 = x_{\rm min}$ and $x_N=x_{\rm max}$.
\end{itemize}

Gauss quadrature is the best possible choice in terms of degree (it is not possible to find a quadrature with $\delta >1$). However, the other quadratures have the property that the boundaries of the interval coincides with collocation points, which can be useful, to enforce boundary conditions for example. In all the following, we will use the Gauss-Lobatto quadrature.

\subsubsection{Interpolation}
If one applies the Gauss quadratures to approximate the coefficient of the expansion, one obtains 
\be
\label{tilde_u}
\tilde{u}_n = \fraction{1}{\gamma_n}
\sum_{j=0}^N u\l(x_j\r) p_n\l(x_j\r) w_j \quad {\rm with} \quad 
\gamma_n = \sum_{j=0}^N p_n^2 \l(x_j\r) w_j.
\ee
Let us precise that this is not exact in the sense that $\hat{u}_n \not= \tilde{u}_n$. However, the computation of $\tilde{u}$ only requires to evaluate $u$ at the $N+1$ collocation points. The {\em interpolant} of $u$ is then defined as the following polynomial 
\be
\label{interpol}
I_N u = \sum_{n=0}^N \tilde{u}_n p_n\l(x\r).
\ee
The difference between $I_nu$ and $P_nu$ is called the {\em aliasing error}. The interpolant of $u$ is the spectral approximate of $u$ and one can show that it is the only polynomials of degree $N$ that coincides with $u$ at each collocation point : 
\be
\l[I_nu\r]\l(x_i\r) = u\l(x_i\r) \quad \forall i\leq N.
\ee

\begin{figure}
\centerline{
\includegraphics[height=5.5cm]{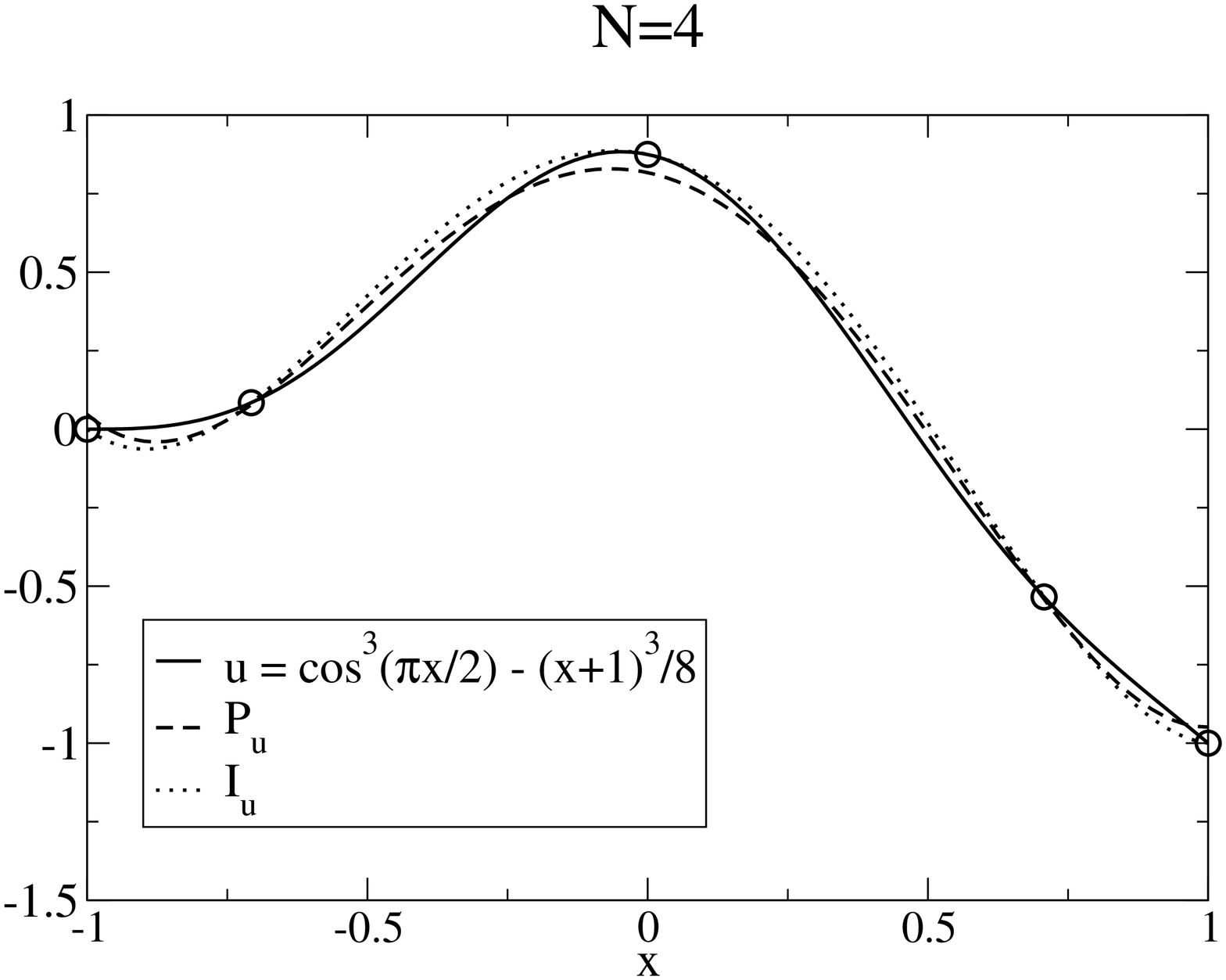}
\includegraphics[height=5.5cm]{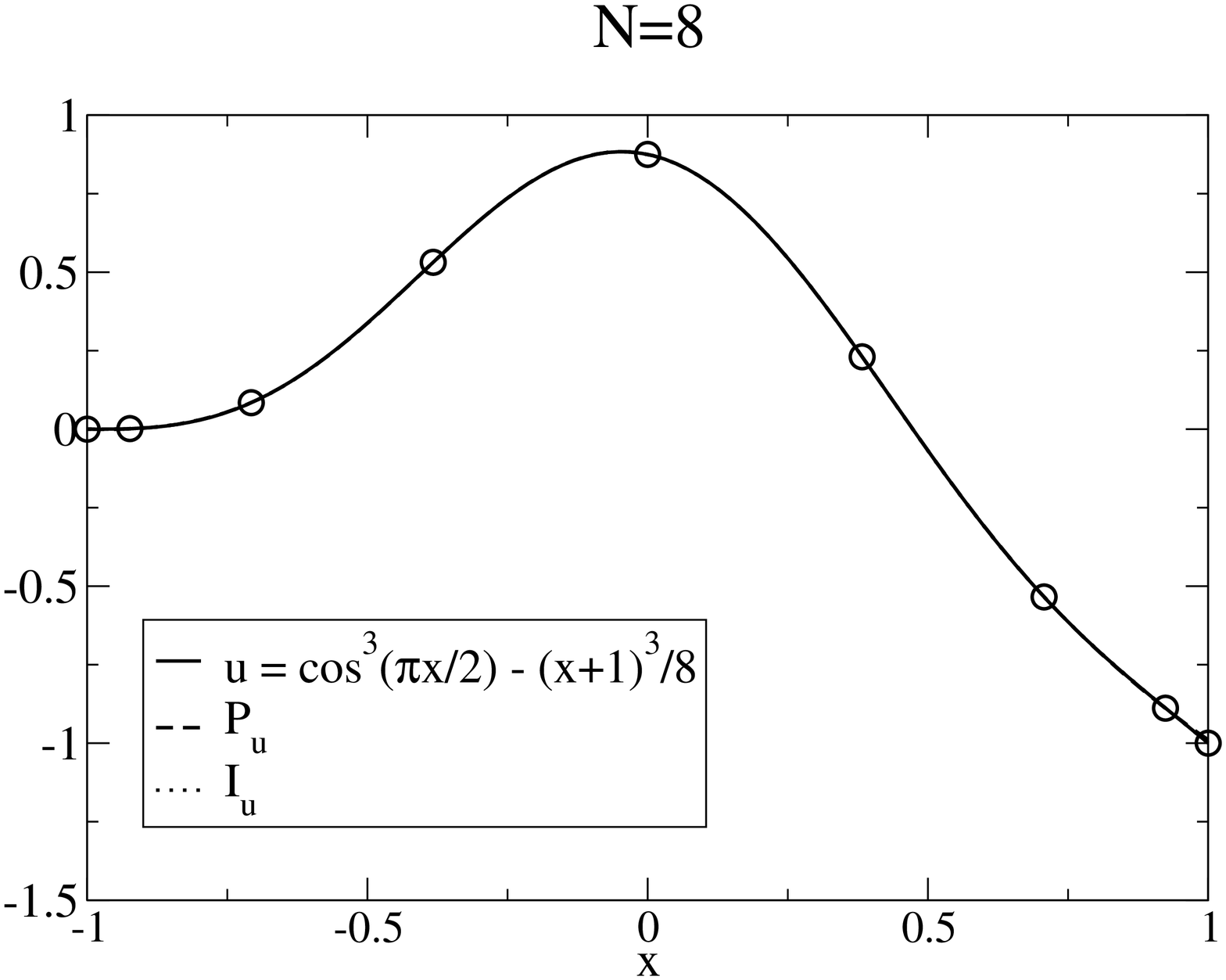}}
\caption{\label{f:interpol}
Same as Fig. \ref{f:proj} with also the interpolant of $u$. The collocation points are denoted by the circles.
}
\end{figure}

Figure \ref{f:interpol} shows the same function as Fig.\ref{f:proj} but the interpolant is also plotted. One can see that, indeed $I_Nu$ coincides with $u$ at the collocation points that are indicated by the circles. Once again, even with as few points as $N=8$, no difference can be seen between the various functions.

\begin{figure}
\centerline{\includegraphics[height=8cm]{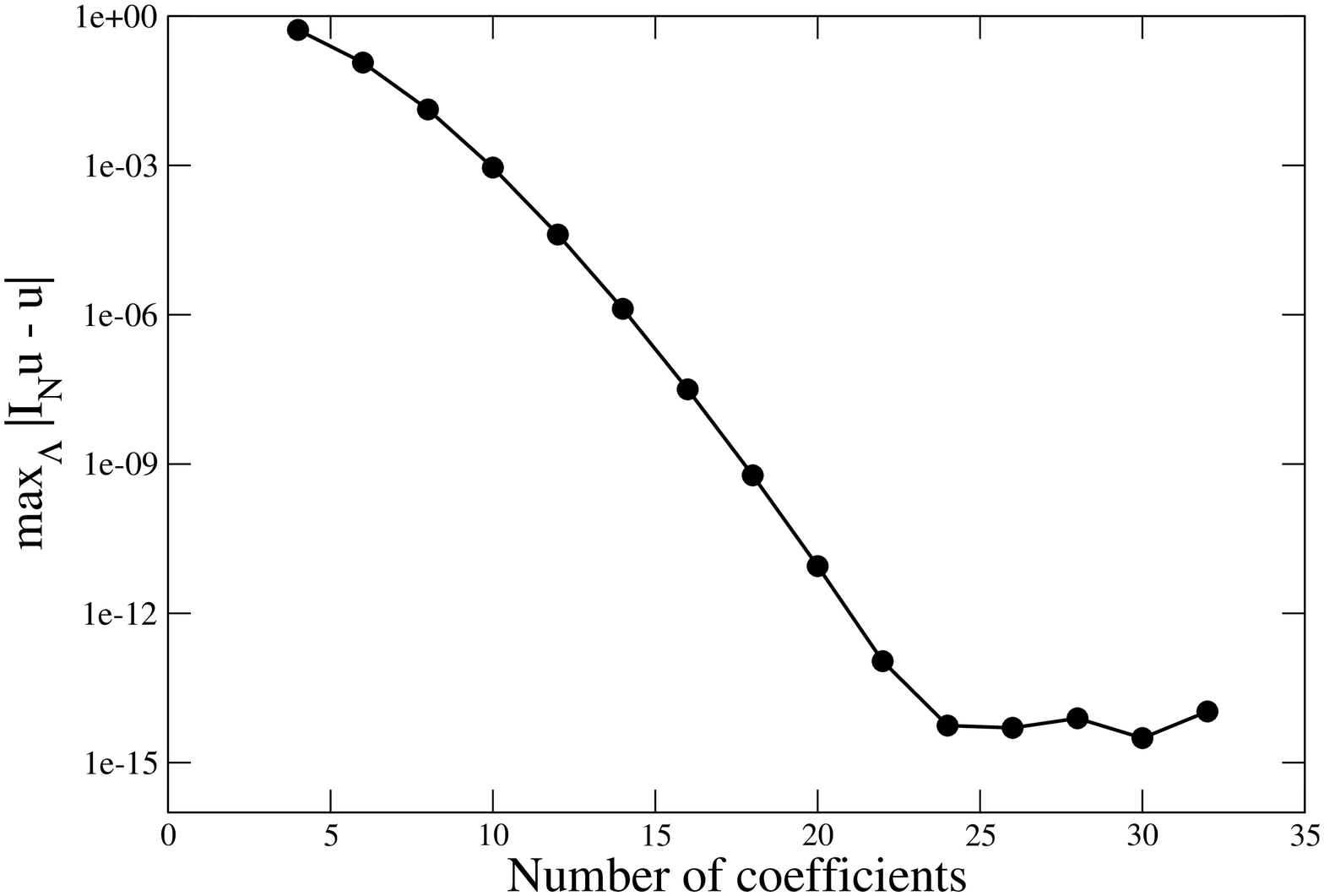}}
\caption{\label{f:conv_interpol}
Maximum difference between $I_Nu$ and $u$ as a function of the degree of the approximation $N$.
}
\end{figure}

Figure \ref{f:conv_interpol} shows the maximum difference between $I_N u$ and $u$ on $\Lambda$, as a function of the degree of the approximation $N$. We can observe the very general feature of spectral methods that the error decreases exponentially, until one reaches the machine accuracy (here $10^{-14}$, the computation being done in double precision). This very fast convergence explains why spectral methods are so efficient, especially compared to finite difference ones, where the error follows only a power-law in terms of $N$. We will later be more quantitative about the convergence properties of the spectral approximation.

Let us note that a function $u$ can be described either by its value at each collocation point $u\l(x_i\r)$ or by 
the coefficients of the interpolant $\tilde{u}_i$. If the values at collocation points are known one is working in the {\em configuration space} and in the {\em coefficient space} if $u$ is given in terms of its coefficients.
There is a bijection between the two descriptions and one simply goes from one space to another by using : 
\begin{itemize}
\item $\tilde{u}_n = \fraction{1}{\gamma_n}\sum_{j=0}^N u\l(x_j\r) p_n\l(x_j\r) w_j$ (configuration $\rightarrow$ coefficient)
\item $u\l(x_n\r) = \displaystyle\sum_{j=0}^N \tilde{u}_j p_j\l(x_n\r)$ \quad \quad\quad (coefficient $\rightarrow$ configuration)
\end{itemize}

\begin{figure}
\centerline{
\includegraphics[height=5.5cm]{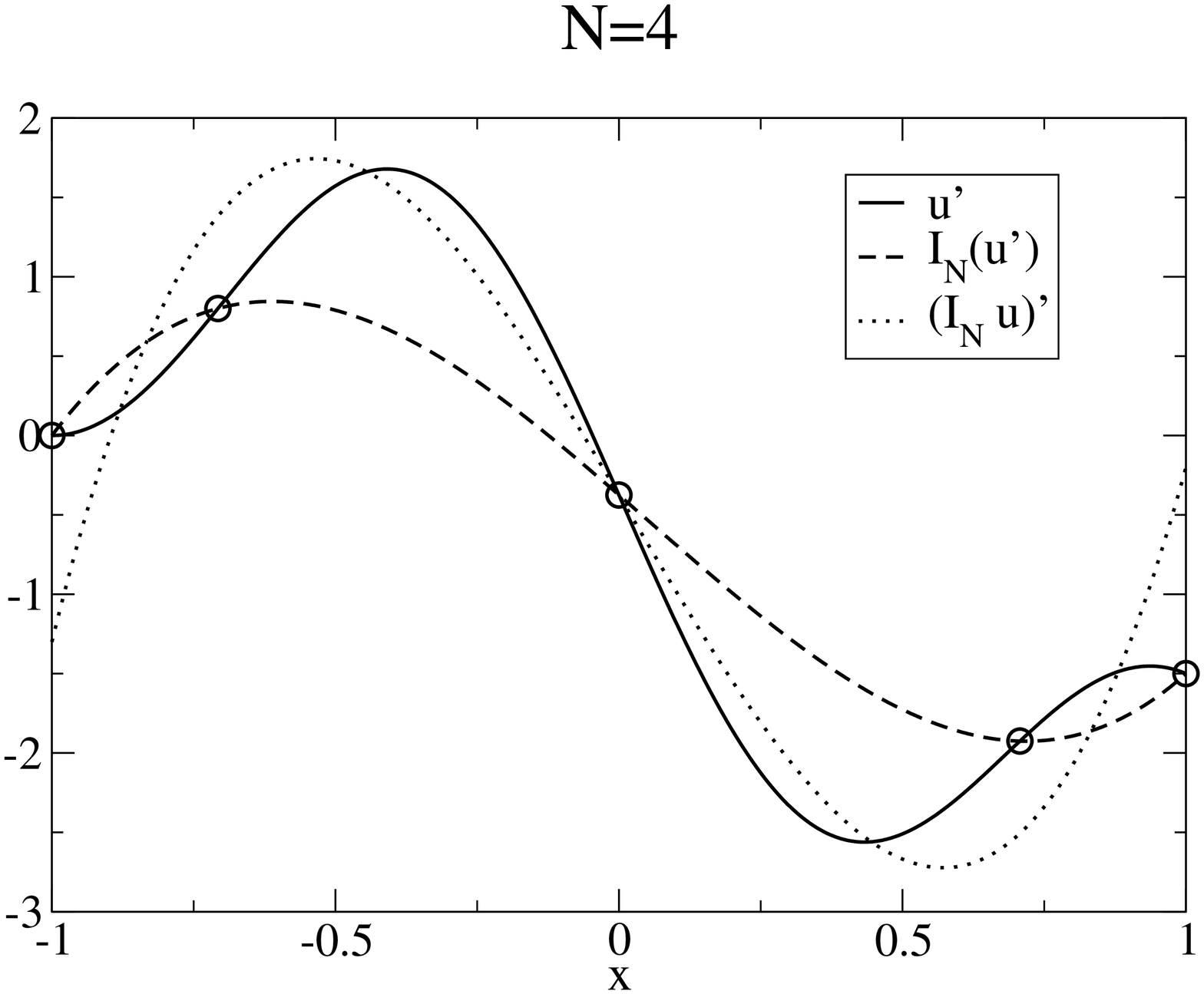}
\includegraphics[height=5.5cm]{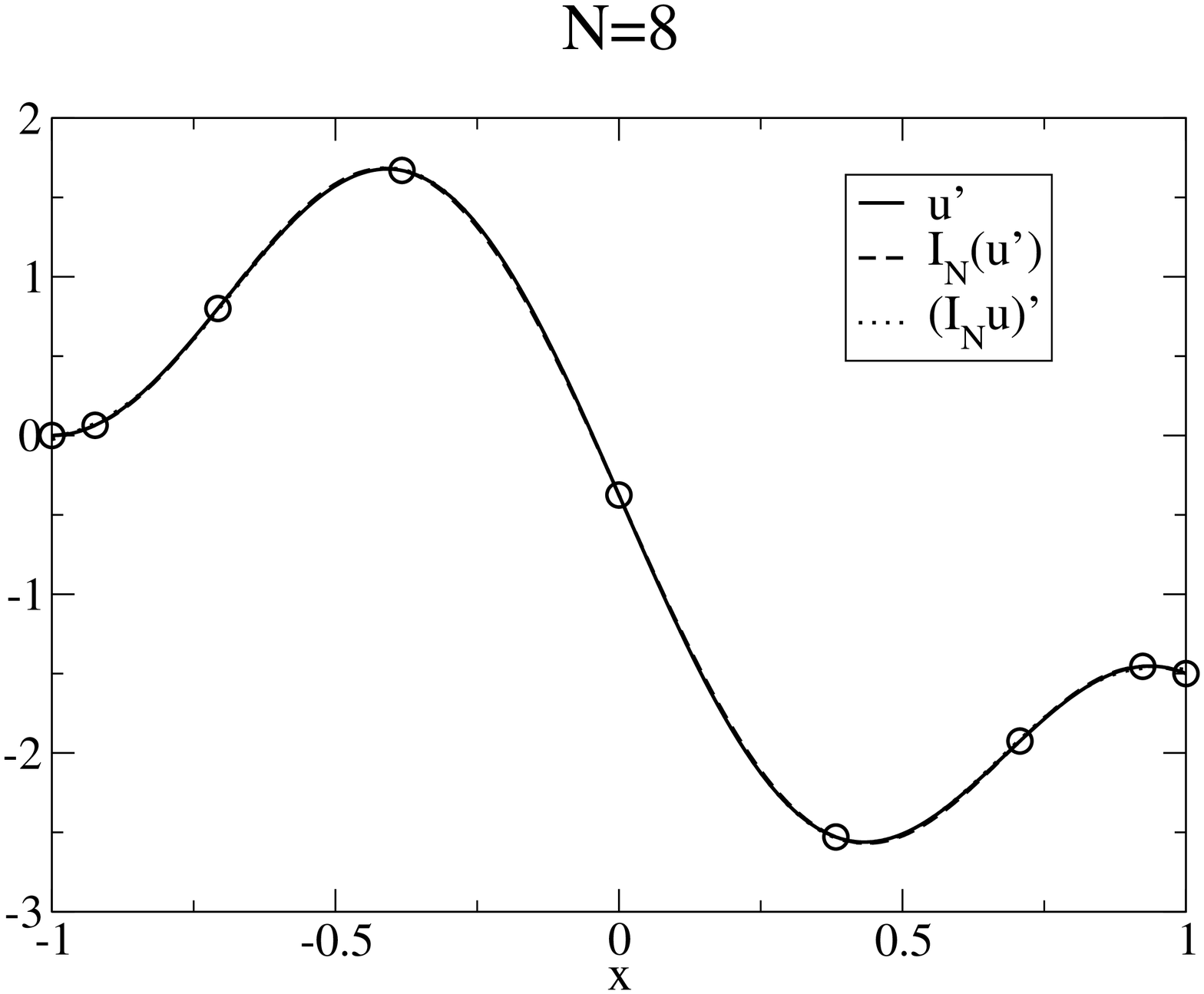}}
\caption{\label{f:der}
First derivative of $u= \cos^3\l(\pi x/2\r)-\l(x+1\r)^3/8$, interpolant of the derivative and derivative of 
the interpolant, for $N=4$ and $N=8$.
}
\end{figure}

Depending on the operation one has to perform, one choice of space is usually more suited than the other. For instance, let us assume that one wants to compute the derivative of $u$. This is easily done if $u$ is known in the coefficient space. Indeed, one can simply approximate $u'$ by the derivative of the interpolant : 
\be
u'\l(x\r) \approx \l[I_N u\r]' = \sum_{n=0}^N \tilde{u}_n p_n'\l(x\r).
\ee
Such an approximation only requires the knowledge of the coefficients of $u$ and how the basis polynomials are derived. Let us note that the obtained polynomial, even if it is a good approximation of $u'$, is not the interpolant of $u'$. In other terms, the interpolation and the derivation are two operations that do not commute:~$\l(I_Nu\r)' \not= I_N \l(u'\r)$. This is clearly illustrated on Fig. \ref{f:der} where the derivative of 
$u= \cos^3\l(\pi x/2\r)-\l(x+1\r)^3/8$ is plotted along with the functions $\l(I_Nu\r)'$ and $I_N \l(u'\r)$. 
In particular, one can note that the functions used to represent $u'$, i.e. $\l(I_Nu\r)'$ does not coincide with $u'$ at the collocation points. However, even with $N=8$ only, the three functions can not be distinguished by eye.

\begin{figure}
\centerline{
\includegraphics[height=8cm]{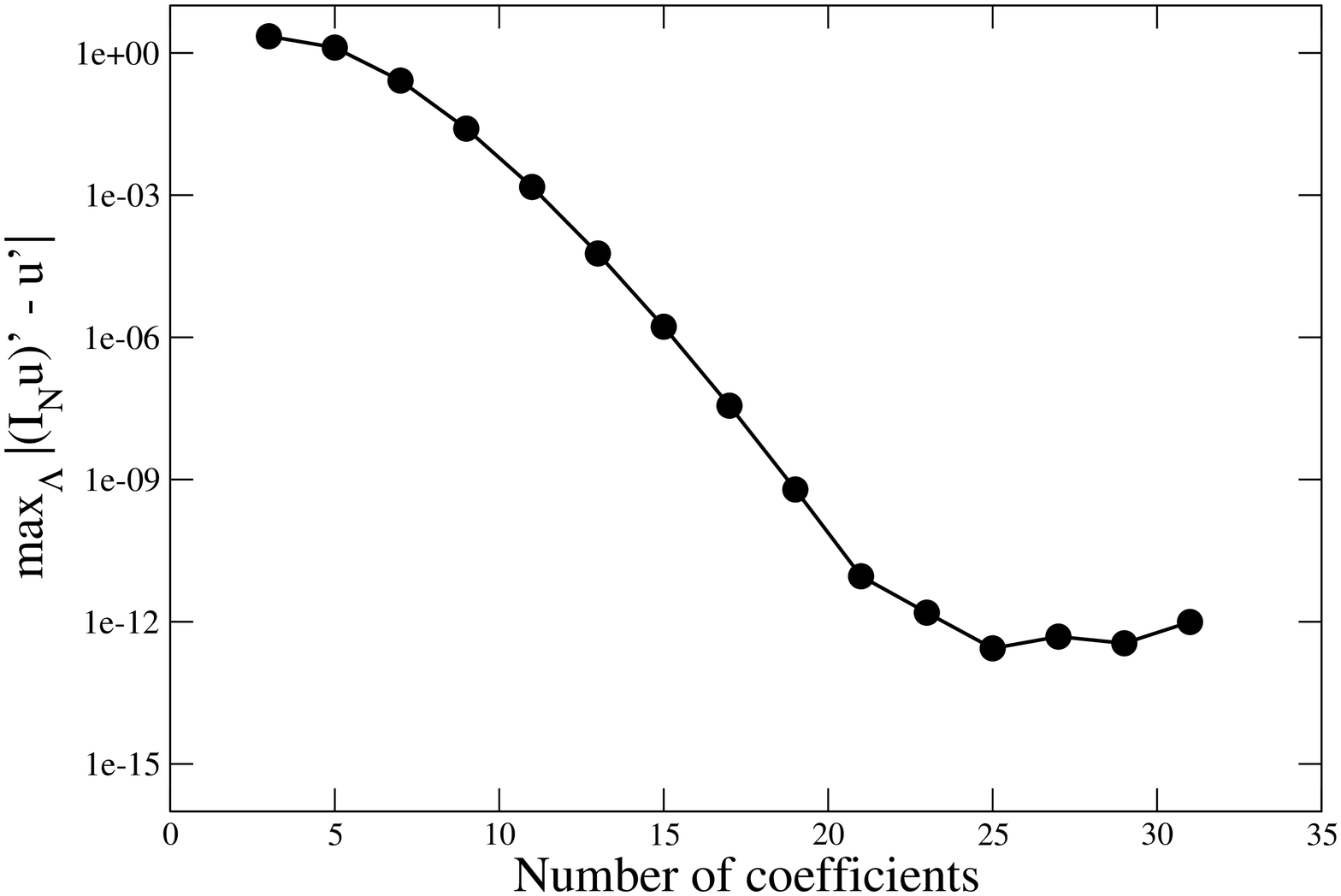}}
\caption{\label{f:conv_der}
Maximum difference between $\l(I_Nu\r)'$ and $u'$ as a function of the degree of the approximation $N$.
}
\end{figure}

The maximum difference between $u'$ and $\l(I_Nu\r)'$, as a function of $N$, is shown on Fig. \ref{f:conv_der}. Once, again, the convergence is exponential.

\subsection{Usual families of polynomials}\label{poly}
In this section, we briefly present some basic properties of Legendre and Chebyshev polynomials. Those two sets are the usual choice for non periodic problems.
\subsubsection{Legendre polynomials}
The Legendre polynomials, denoted by $P_n$, constitute a family of orthogonal polynomials on $\l[-1, 1\r]$ with a measure $w=1$. The fact that the measure is so simple is one of the main advantage of working with Legendre polynomials, especially from the analytical point of view.

The scalar product of two $P_n$ is given by :
\be
\int_{-1}^1 P_n P_m {\rm d}x = \frac{2}{2n+1} \delta_{mn}.
\ee

\begin{figure}
\centerline{
\includegraphics[height=8cm]{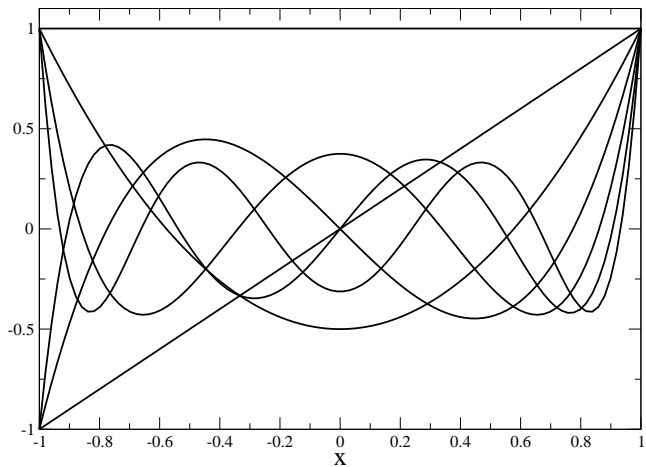}}
\caption{\label{f:legendre}
The first Legendre polynomials, from $P_0$ to $P_6$.
}
\end{figure}

The successive polynomials can be constructed by recurrence. Indeed given that $P_0 = 1$ and $P_1 = x$, all the $P_n$ can be obtained by using 
\be
\label{rec_leg}
\l(n+1\r) P_{n+1}\l(x\r) = \l(2n+1\r) x P_n\l(x\r) - n P_{n-1}\l(x\r).
\ee
It is then easy to see that the Legendre polynomials have the following simple properties : i) $P_n$ has the same parity as $n$. ii) $P_n$ is of degree $n$. iii) $P_n\l(\pm 1\r) = \l(\pm 1\r)^{n}$. iv) $P_n$ has exactly $n$ zero on
$\l[-1,1\r]$. The first polynomials are plotted on Fig. \ref{f:legendre}.

The values of the weights and collocation points can be written for the three usual quadratures :
\begin{itemize}
\item Legendre-Gauss : $x_i$ are the nodes of $P_{N+1}$ and $w_i = \fraction{2}{\l(1-x_i^2\r)
\l[P'_{N+1} \l(x_i\r)\r]^2}$.
\item Legendre-Gauss-Radau : $x_0=-1$ and the $x_i$ are the nodes of $P_N + P_{N+1}$. The weights are given by 
$w_0 = \fraction{2}{\l(N+1\r)^2}$ and $w_i = \fraction{1}{\l(N+1\r)^2}$.
\item Legendre-Gauss-Lobatto : $x_0=-1$, $x_N=-1$ and $x_i$ are the nodes of $P'_N$. The weights are 
$w_i = \fraction{2}{N\l(N+1\r)}\fraction{1}{\l[P_N\l(x_i\r)\r]^2}$.
\end{itemize}
From the above formula, one can see that the position of the collocation points are not analytical and they have to be computed numerically, which is one of the main shortcoming of Legendre polynomials.

One can also derive the action of some linear operation in the coefficient space. Consider a function $f$ given by its coefficients : $f=\sum_{n=0}^N a_n P_n\l(x\r)$ and $H$ be a linear operator acting on $f$ so that $Hf = \sum_{n=0}^N b_n P_n\l(x\r)$. The relation between the $a_n$ and $b_n$ can be explicitly  written in some cases. For example :
\begin{itemize}
\item If $H$ is the multiplication by $x$ then : 
\be
b_n = \frac{n}{2n-1}a_{n-1} + \frac{n+1}{2n+3} a_{n+1}\quad \l(n \geq 1\r).
\ee
\item If $H$ is the derivation : 
\be
b_n = \l(2n+1\r) \sum_{p=n+1, p+n \, {\rm odd}}^N a_p.
\ee
\item If $H$ is the second derivation : 
\be
b_n = \l(n+1/2\r) \sum_{p=n+2, p+n \, {\rm even}}^N \l[p\l(p+1\r) - n \l(n+1\r)\r] a_p.
\ee
\end{itemize}
Those kind of relations enable to represent the action of $H$ as a matrix acting on the vector of the $a_n$, the product being the coefficients of $Hf$, i.e. the $b_n$.

\subsubsection{Chebyshev polynomials}
The Chebyshev polynomials $T_n$ are an orthogonal set on $\l[-1, 1\r]$ for the measure 
$w = \fraction{1}{\sqrt{1-x^2}}$. More precisely one has
\be
\int_{-1}^1 \frac{T_n T_m}{\sqrt{1-x^2}} {\rm d}x = \frac{\pi}{2}\l(1+\delta_{0n}\r) \delta_{mn}.
\ee

\begin{figure}
\centerline{
\includegraphics[height=8cm]{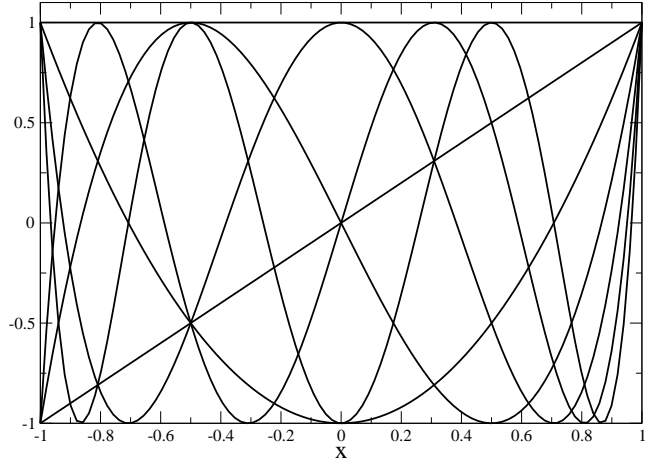}}
\caption{\label{f:cheby}
The first Chebyshev polynomials, from $T_0$ to $T_6$.
}
\end{figure}

Chebyshev polynomials can be computed by knowing that $T_0 = 1$, $T_1 = x$ and by making use of the recurrence : 
\be
\label{rec_cheb}
T_{n+1} \l(x\r) = 2x T_n\l(x\r) - T_{n-1}\l(x\r).
\ee
It follows that i) $T_n$ has the same parity as $n$. ii) $T_n$ is of degree $n$. iii) $T_n\l(\pm 1\r) = \l(\pm 1\r)^{n}$. iv) $T_n$ has exactly $n$ zero on
$\l[-1,1\r]$. The first polynomials are plotted on Fig. \ref{f:cheby}.

The weights and collocation points associated with Chebyshev polynomials can be computed : 
\begin{itemize}
\item Chebyshev-Gauss : $x_i=\cos\fraction{\l(2i+1\r)\pi}{2N+2}$ and $w_i = \fraction{\pi}{N+1}$.
\item Chebyshev-Gauss-Radau : $x_i = \cos\fraction{2\pi i }{2N+1}$. The weights are $w_0= \fraction{\pi}{2N+1}$ and 
$w_i =  \fraction{2\pi}{2N+1}$
\item Chebyshev-Gauss-Lobatto :  $x_i = \cos\fraction{\pi i }{N}$. The weights are $w_0= w_N= \fraction{\pi}{2N}$ and 
$w_i =  \fraction{\pi}{N}$.
\end{itemize}
Contrary to Legendre polynomials, the position of the collocation points are completely analytical which can somewhat simplify the computational task. In all the following, Chebyshev polynomials
are used, except when otherwise stated.

Once again, one can find the relation between the coefficients $a_n$ of a function $f$ and the coefficients $b_n$ of $Hf$, where $H$ is a linear operator. For example : 
\begin{itemize}
\item If $H$ is the multiplication by $x$ then : 
\be
b_n = \frac{1}{2}\l[\l(1+\delta_{0\, n-1}\r) a_{n-1} + a_{n+1}\r] \quad \l(n \geq 1\r).
\ee
\item If $H$ is the derivation : 
\be
b_n = \frac{2}{\l(1+\delta_{0\, n}\r)} \sum_{p=n+1, p+n \, {\rm odd}}^N pa_p.
\ee
\item If $H$ is the second derivation : 
\be
b_n = \frac{1}{\l(1+\delta_{0\, n}\r)} \sum_{p=n+2, p+n \, {\rm even}}^N p\l(p^2-n^2\r) a_p.
\ee
\end{itemize}

\subsection{Convergence properties}
As already seen, for sufficiently regular functions, the difference between $u$ and its interpolant $I_Nu$ goes to zero exponentially. This statement can be made more precise. Let us consider a ${\mathcal C}^m$ function $u$. Upper bounds on the difference between $u$ and $I_Nu$ can be found for various norms and choice of polynomials.
\begin{itemize}
\item For Legendre : 
\be
\l\| I_Nu-u \r\|_{L^2} \leq \frac{C}{N^{m-1/2}}\sum_{k=0}^m \l\|u^{\l(k\r)}\r\|_{L^2}.
\ee
\item For Chebyshev : 
\be
\l\| I_Nu-u \r\|_{L_w^2} \leq \frac{C}{N^m}\sum_{k=0}^m \l\|u^{\l(k\r)}\r\|_{L_w^2}.
\ee
\be
\l\| I_Nu-u \r\|_\infty \leq \frac{C}{N^{m-1/2}}\sum_{k=0}^m \l\|u^{\l(k\r)}\r\|_{L_w^2}.
\ee
\end{itemize}

Without going into to much details, one can note that the errors decay faster than some power of $N$. The upper bounds decay faster for more regular function (i.e. for higher $m$). An interesting case is the one of a 
${\mathcal C}^\infty$ function. In such a case, the error decays faster than any power of $N$ and thus like an exponential. One talks of an {\em evanescent error} and this is the case previously observed. 

\begin{figure}
\centerline{
\includegraphics[height=8cm]{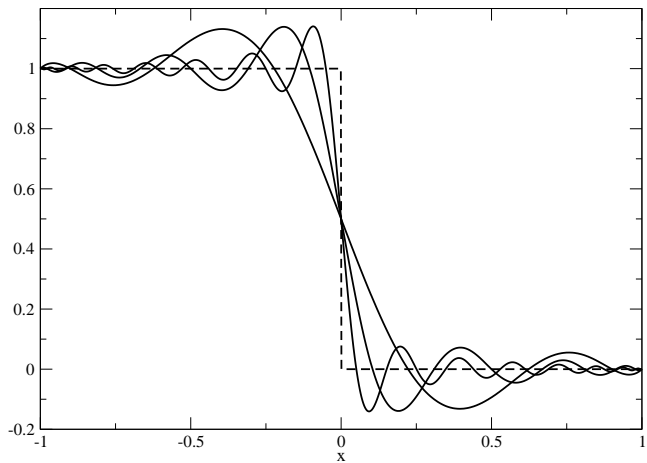}}
\caption{\label{f:gibbs}
Step function and some interpolants for $N=7$, $N=15$ and $N=31$. As $N$ increases the maximum difference stays constant and the number of oscillations increases.
}
\end{figure}

\begin{figure}
\centerline{
\includegraphics[height=8cm]{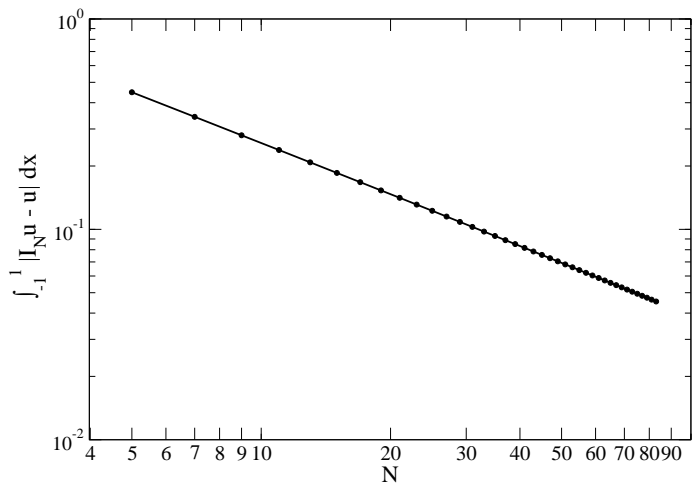}}
\caption{\label{f:gibbs_int}
$\int_{-1}^{1} \l|I_N u - u \r| {\rm d}x$ as a function of $N$, $u$ being a step function. The convergence obeys a power-law which decays slower than $N^{-1}$.
}
\end{figure}

A limit case can be provided by a discontinuous function. Consider a step function $u$ such that $u\l(x<0\r) = 1$ and $u\l(x>0\r) = 0$. If one tries to interpolate this function the convergence theorems previously exposed can not ensure convergence. This reflects on Fig. \ref{f:gibbs} where the step function is presented with interpolants for various number of points. As the number of collocation points increases, the maximum difference between $u$ and $I_Nu$ stays constant (the amplitude of the oscillations can be seen to be constant). However, more and more oscillations are present, as the Chebyshev polynomials do their best to approximate the discontinuous step function. In a sense, the approximation is still better and better, the support of the difference being smaller and smaller. In other words, the integrated error : $\displaystyle\int_{-1}^{1} \l|I_N u - u \r| {\rm d}x$ still goes to zero when $N$ increases, as can be seen on Fig. \ref{f:gibbs_int}. However, this convergence follows only a very slow-decaying power-law.

The simple example of the step function is a very general feature of the so-called {\em Gibbs phenomenon}, which occurs anytime one tries to interpolate a function that is not ${\mathcal C}^\infty$. When this happens, the error is no longer evanescent and only converges as a power-law. In order to maintain evanescent errors, one should try to avoid any Gibbs phenomenon. In some cases, as will be seen later, this can be achieved by using a multi-domain decomposition of space.

\section{Differential equation solvers}\label{s:oned}
\subsection{The weighted residual method}
The type of problems with are concerned with in the section are ordinary differential equations, in one dimension only, on a bounded domain at the boundaries of which some conditions are enforced on the solution. Mathematically, one considers the following system :
\bea
\label{e:bulk}
Lu\l(x\r) = S\l(x\r) && \quad {\rm for} \quad x\in U\\
\label{e:boundary}
Bu\l(y\r) = 0 && \quad {\rm for} \quad y \in\partial U
\eea
where $L$ and $B$ are linear differential operators. A function $u$ is then an admissible solution of this system, if and only if i) it satisfies Eq. (\ref{e:boundary}) ``exactly'' (i.e. to machine accuracy) ii) it makes the residual $R \equiv Lu - S$ small. In order to quantify what this ``small'' means, the weighted residual method relies on $N+1$ tests functions $\xi_n$ and one asks that the scalar product of $R$ with those functions is exactly zero :
\be
\label{e:residu}
\l(\xi_k, R\r) = 0, \quad \forall\, k \leq N.
\ee
Of course as $N$ increases the obtained solution is closer and closer to the real one. Depending on the choice of spectral basis and of test functions, one can generate various different types of spectral solvers. In the following, three of the most used ones are presented and applied to a simple case.

\subsection{A test problem}
We propose to solve the equation : 
\be
\frac{{\rm d}^2u}{{\rm d}x^2} - 4 \frac{{\rm d}u}{{\rm d}x} + 4u = \exp\l(x\r) + C,
\ee
with $x \in\l[-1,1\r]$ and $C= -4e/\l(1+e^2\r)$. As boundary conditions, we simply ask that the solution is zero at the boundaries : 
\be
\label{e:bound_example}
u\l(-1\r) = 0 \quad {\rm and} \quad u\l(1\r) = 0.
\ee

Under those conditions, the solution is unique and analytical : 
\be
u_{\rm sol}\l(x\r) = \exp\l(x\r) - \frac{\sinh \l(1\r)}{\sinh \l(2\r)} \exp\l(2x\r) + \frac{C}{4}.
\ee
Let us note that this solution is not a polynomial.

With our notations the linear operator $L$ is $\fraction{{\rm d}^2}{{\rm d}x^2} - 4 \frac{{\rm d}}{{\rm d}x} + 4 
{\rm Id}$. Using the elementary linear operations seen in Sec.\ref{poly}, one can construct the matrix representation of $L$, which will come handy in the implementation of the various solvers. Let us recall that if $u = \displaystyle\sum_{i=0}^N \tilde{u}_i T_i\l(x\r)$ then 
$Lu = \displaystyle\sum_{i=0}^N \displaystyle\sum_{j=0}^N L_{ij} \tilde{u}_j T_i \l(x\r)$.

In this particular case, and for $N=4$, one finds :
\be
L_{ij} = \l( \begin{tabular}{ccccc}
4  &  -4 &  4 &  -12  & 32 \\
0  &   4 &-16 &   24  & -32\\
0  &  0  &  4 &   -24 & 48\\
0  &  0  &  0 &  4    & -32 \\
0  &  0  &  0 &  0    &   4 \\
\end{tabular}\r).
\ee

\subsection{The Tau-method}\label{s:tau}
In this method, the test functions $\xi_n$ are chosen to be the same as the spectral functions of decomposition. Let us assume that one uses Chebyshev polynomials $T_i$. The residual equations (\ref{e:residu}) are then : 
\be
\l(T_n, Lu-S\r) = 0 \quad \forall \, n \leq N.
\ee
If one uses the definition of the matrix $L_{ij}$, this set of equations can be written as 
\be
\sum_{j=0}^N L_{nj} \tilde{u}_j = \tilde{s}_n \quad \forall \, n \leq N
\ee
where the $\tilde{s}_n$ are the spectral coefficients of the source $S$. 

However, due to the presence of homogeneous solutions of $L$, this set of $N~+~1$ equations is degenerate and one must impose the boundary conditions before solving it. In the Tau-method, the boundary conditions are enforced as additional equations. In our particular case they can be written in a very straightforward manner :
\bea
\label{e:bound_eq}
u\l(x=-1\r) = 0 &\Longrightarrow& \sum_{j=0}^N \l(-1\r)^j \tilde{u}_j = 0 \\
\nonumber
u\l(x=+1\r) = 0 &\Longrightarrow& \sum_{j=0}^N \tilde{u}_j = 0.
\eea
One has then to relax the two {\em last} residual equations and to replace them by the two boundary conditions in order to get an invertible system for which the unknowns are the $\tilde{u}_n$. Relaxing the last two equations is not an issue. Indeed, if the functions are regular enough, the coefficients are rapidly decreasing and thus the obtained solution should converge nicely to the exact solution. 

In our example, and for $N=4$ the system reads as : 
\be
\l( \begin{tabular}{ccccc}
4  &  -4 &  4 &  -12  & 32 \\
0  &   4 &-16 &   24  & -32\\
0  &  0  &  4 &   -24 & 48\\
1  &  -1  &  1 &  -1    & 1 \\
1  &  1  &  1 &  1    &   1 \\
\end{tabular}\r) 
\l( \begin{tabular}{c}
$\tilde{u_0}$ \\
$\tilde{u_1}$ \\
$\tilde{u_2}$ \\
$\tilde{u_3}$ \\
$\tilde{u_4}$ \\
\end{tabular}\r) = 
\l( \begin{tabular}{c}
-0.03 \\
1.13 \\
0.27 \\
0 \\
0 \\
\end{tabular}\r).
\ee
Once inverted, the coefficients of the solution can be found
$\tilde{u_0}\simeq 0.146 ; \tilde{u_1}\simeq 0.079 ; \tilde{u_2} \simeq -0.122 ; \tilde{u_3}\simeq -0.079 ; \tilde{u_4}\simeq -0.024$ (for $N=4$). 

\begin{figure}
\centerline{
\includegraphics[height=5.5cm]{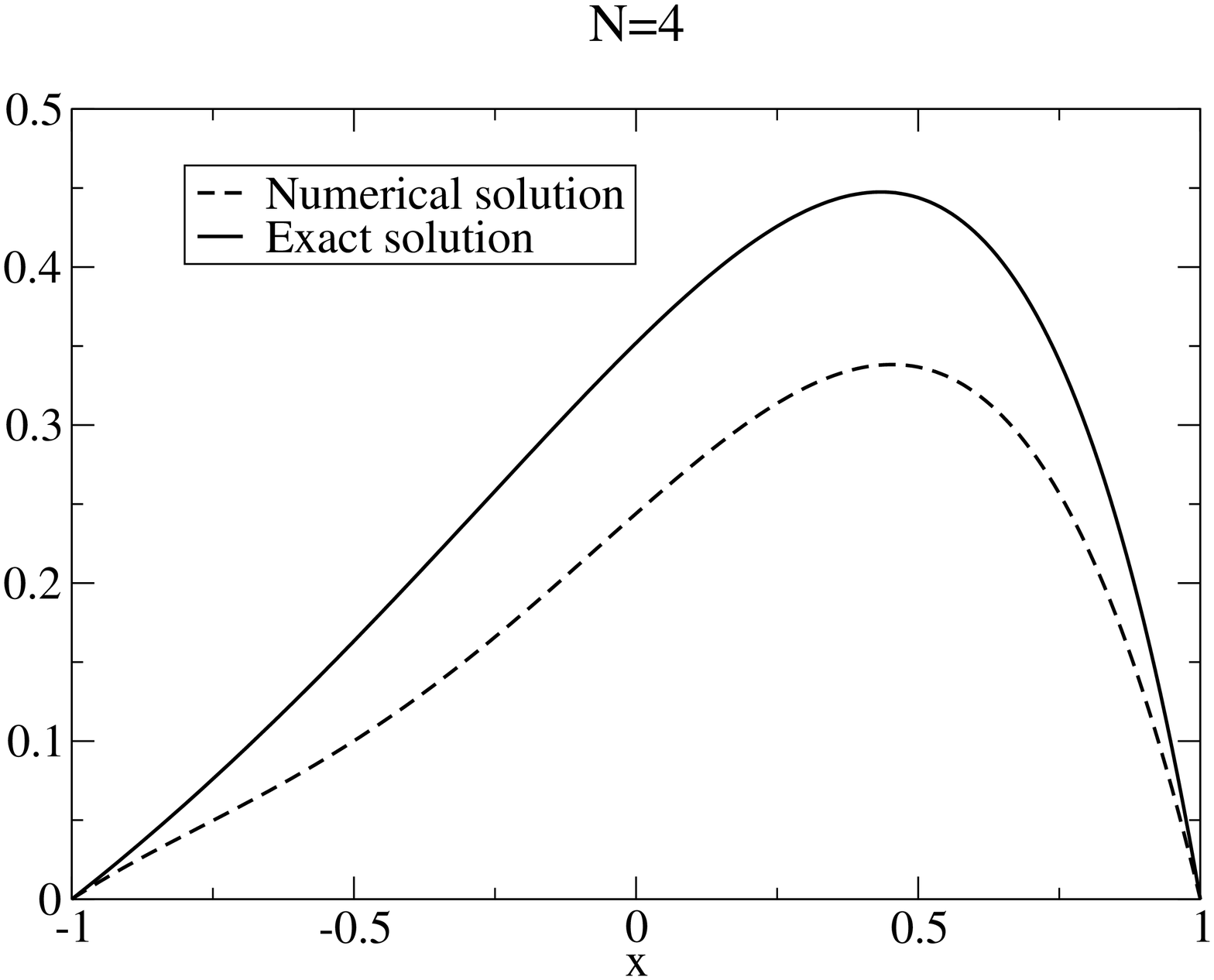}
\includegraphics[height=5.5cm]{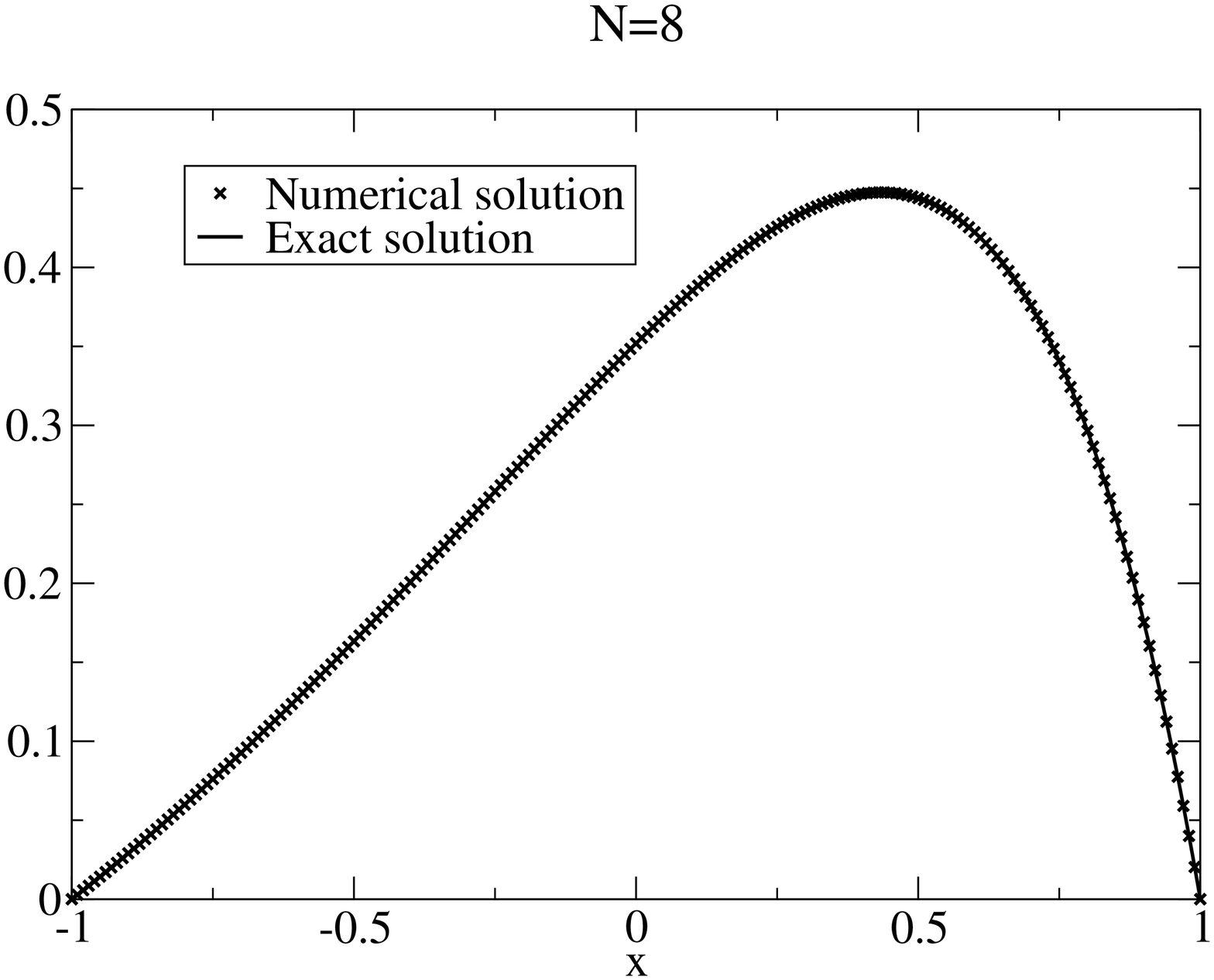}}
\caption{\label{f:tau}
Exact solution (solid line) and numerical one (dashed line and cross symbols), for the Tau-method, for $N=4$ and $N=8$.
}
\end{figure}

The numerical solution obtained is compared to the exact one on Fig. \ref{f:tau}. If the difference is easily seen for $N=4$, even with as few points as $N=8$, there is not difference noticeable by eye between the two curves. One can also note, as expected when using the weighted residual method, that, for all $N$, the numerical solution ''exactly'' fulfills the boundary conditions.

\subsection{The collocation method}
In this method, the test functions are no longer the functions of the spectral expansion, but are chosen to be zero but at one collocation point $\xi_n = \delta \l(x-x_n\r)$. The residual equations
(\ref{e:residu}) are then equivalent to : 
\be
Lu\l(x_n\r) = S\l(x_n\r) \quad \forall n \leq N.
\ee

Given the expression of the coefficients of $Lu$ in terms of the matrix $L_{ij}$, one can write those equations as a system for the unknowns $\tilde{u}_n$ : 
\be
\label{e:syst_colloc}
\sum_{i=0}^N \sum_{j=0}^N L_{ij} T_i \l(x_n\r) \tilde{u}_j  = S\l(x_n\r) \quad \forall n\leq N.
\ee
As for the Tau-method, this system is degenerate and must be supplied with the boundary conditions. One again, they are enforced by two additional equations, which are the same than for the Tau-method (i.e. Eqs (\ref{e:bound_eq})).However, instead of relaxing the last two equations of (\ref{e:syst_colloc}), one has to relax the first and the last ones. Indeed, those are the equations that involved the extremal points, at which the boundary conditions have to be fulfilled. 

For $N=4$, the system for the collocation method is :
\be
\l( \begin{tabular}{ccccc}
1  &  -1 &  1 &  -1  & 1 \\
4  &   -6.83 & 15.3 &   -26.1  & 28\\
4  &  -4  &  0 &   12 & -12\\
4  &  -1.17  &  -7.31 &  2.14    & 28 \\
1  &  1  &  1 &  1    &   1 \\
\end{tabular}\r) 
\l( \begin{tabular}{c}
$\tilde{u_0}$ \\
$\tilde{u_1}$ \\
$\tilde{u_2}$ \\
$\tilde{u_3}$ \\
$\tilde{u_4}$ \\
\end{tabular}\r) = 
\l( \begin{tabular}{c}
0 \\
-0.80 \\
-0.30 \\
0.73 \\
0 \\
\end{tabular}\r)
\ee
and the coefficients of the solution are : 
$\tilde{u_0}\simeq 0.188 ; \tilde{u_1}\simeq 0.089 ; \tilde{u_2} \simeq -0.157 ; \tilde{u_3}\simeq -0.089 ; \tilde{u_4}\simeq -0.031$.

\begin{figure}
 \centerline{
\includegraphics[height=5.5cm]{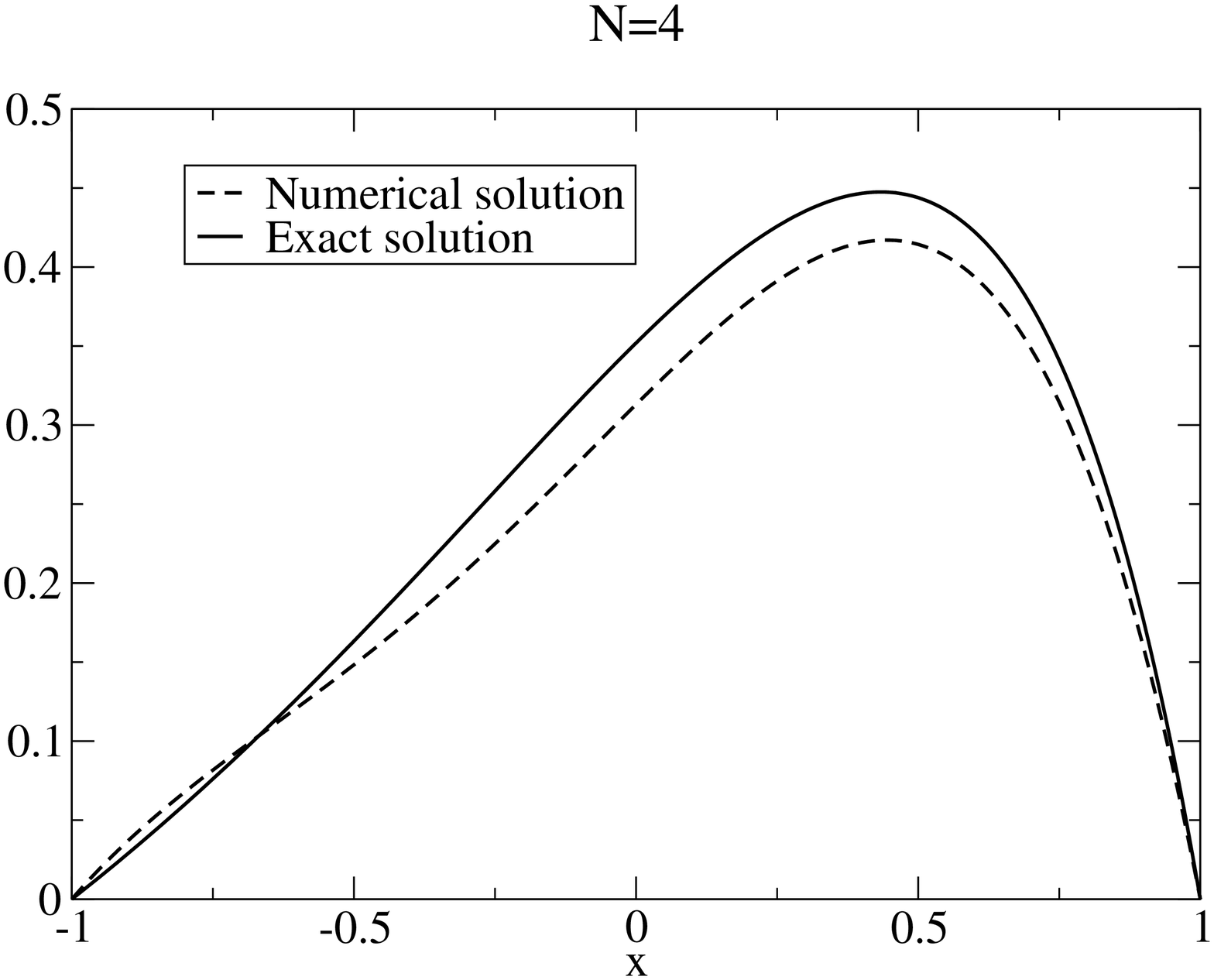}
\includegraphics[height=5.5cm]{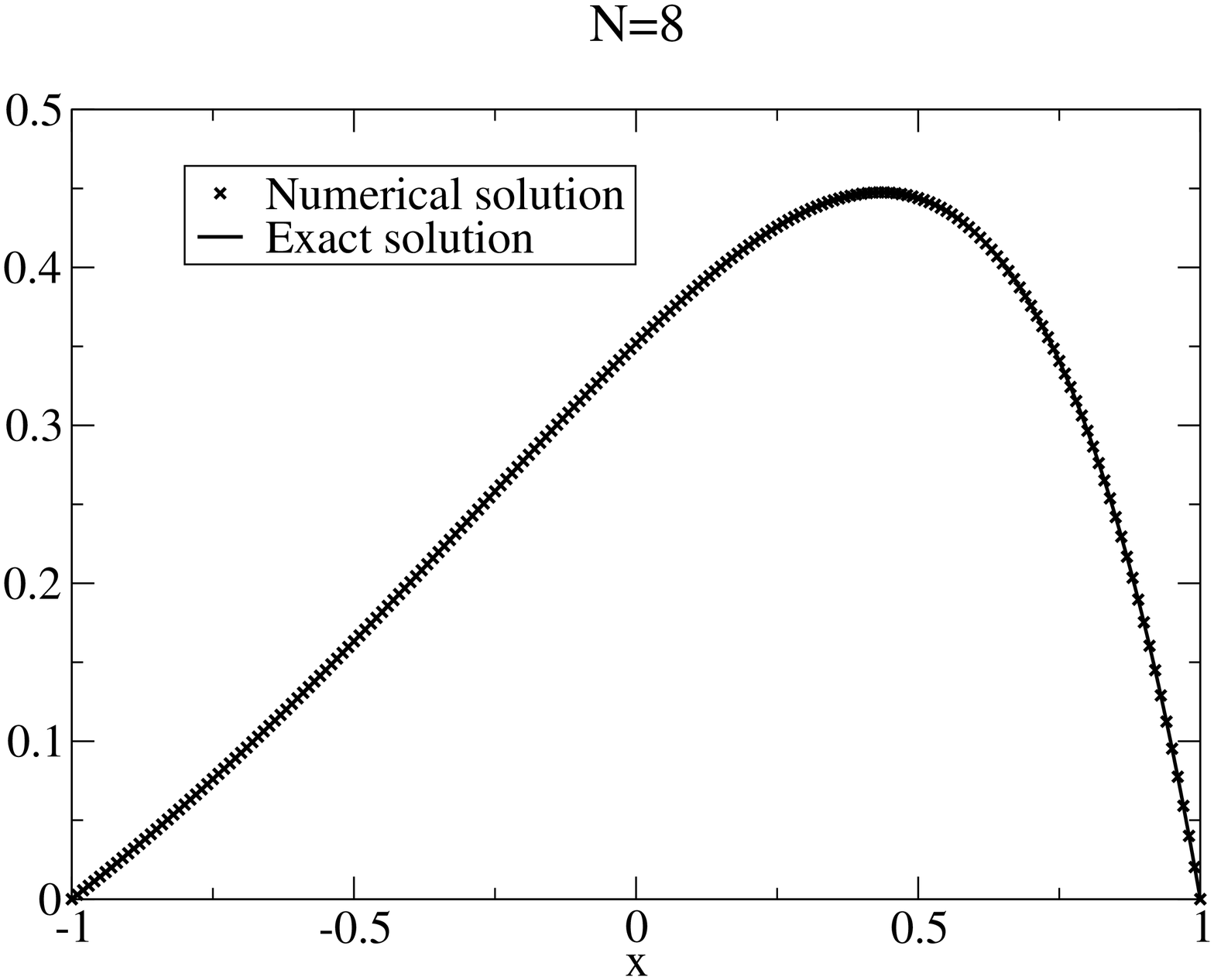}}
\caption{\label{f:colloc}
Exact solution (solid line) and numerical one (dashed line and cross symbols), for the collocation method, for $N=4$ and $N=8$.
}
\end{figure}

Once again, the numerical solution and the exact one are shown on Fig. \ref{f:colloc} and the same properties than for the Tau-method can be observed.

\subsection{The Galerkin method}
The basic idea of the Galerkin method is to expand the solution, not in terms of usual orthogonal polynomials, but on some linear combinations of polynomials that fulfill the boundary conditions. One then talks of {\em Galerkin basis}. The particular choice of basis is of course important and it is rather hard to give a general recipe. However, it is usually better if the Galerkin basis can be easily written in terms of the original basis. 

For the boundary conditions of our example (\ref{e:bound_example}), it is easy to see that the following choice of Galerkin basis $G_i$ is a valid one :
\begin{itemize}
\item $G_{2k}\l(x\r) = T_{2k+2}\l(x\r) - T_0 \l(x\r)$ 
\item $G_{2k+1}\l(x\r) = T_{2k+3}\l(x\r) - T_1 \l(x\r)$.
\end{itemize}

Let us note that, in order to maintain a consistent degree of approximation, one has to consider only $N-1$ Galerkin polynomials (so that the Chebyshev polynomial of higher degree remains $T_N$). This is evident on our example but is a general feature. The reader should also note that the $G_i$, in general, are not orthogonal polynomials.

It is convenient to define the {\em transformation matrix} as the matrix that relates the $G_i$ to the standard polynomials, let us say the $T_i$. This transformation matrix $M$ is not square but a 
$(N+1) \times (N-1)$ matrix and is defined by 
\be
G_i = \sum_{j=0}^N M_{ji} T_j \quad \forall i\leq N-2.
\ee
In our example $M$ is easily constructed and for $N=4$ one gets
\be
M_{ij} = \l( \begin{tabular}{ccc}
-1  &  0 &  -1  \\
0  &   -1 & 0 \\
1 &  0  &  0\\
0  &  1  &  0\\
0  &  0  & 1  \\
\end{tabular}\r).
\ee

As already mentioned, the solution $u$ is seek in terms of the Galerkin basis : 
$u = \displaystyle\sum_{k=0}^{N-2} \tilde{u}^G_k G_k \l(x\r)$. The transformation matrix enables us to get $u$ in terms of $T_i$, and then to use the matrix $L_{ij}$ to get the expression of 
\be
\label{e:lu_galerkin}
Lu\l(x\r) = \sum_{k=0}^{N-2} \tilde{u}^G_k \sum_{i=0}^N\sum_{j=0}^N M_{jk} L_{ij} T_i \l(x\r).
\ee
In order to use the weighted residual method, one has to precise our choice of test functions. In a standard Galerkin method, the $\xi_n$ are chosen to be the Galerkin coefficients so that the residual equation reads $\l(G_n, R\r) \quad \forall n\leq N-2$.

$\l(Lu, G_n\r)$ can be computed by using Eq. (\ref{e:lu_galerkin}) and by expressing $G_n$ in terms of Chebyshev polynomials, using, once again, the transformation matrix.

The situation for the source term is a bit different. Indeed, the source $S$ has no reason at all to verify the same boundary conditions as the solution $u$. So, one has to expand $S$ on the standard basis : $S= \displaystyle\sum_{i=0}^N \tilde{s}_i T_i$. The scalar product $\l(S, G_n\r)$ can then be obtained by using $M$ to go from the $G_i$ basis to the $T_i$ one. 

Putting all the pieces together, the Galerkin system then reads :
\be
\sum_{k=0}^{N-2} \tilde{u}^G_k \sum_{i=0}^N\sum_{j=0}^N M_{in}M_{jk}L_{ij} \l(T_i|T_i\r) = 
\sum_{i=0}^N M_{in} \tilde{s}_i  \l(T_i|T_i\r), \quad \forall n \leq N-2.
\ee

This system is well-posed and can be inverted to get the unknowns, which are the coefficient of $u$ on the Galerkin basis, i.e. the $\tilde{u}^G_i$. Once those are known, the transformation matrix can be used, one last time, to get the solution in terms of the standard spectral basis : 
\be
u\l(x\r) = \sum_{i=0}^N \l(\sum_{n=0}^{N-2} M_{in} \tilde{u}^G_n\r) T_i.
\ee

\begin{figure}
\centerline{
\includegraphics[height=5.5cm]{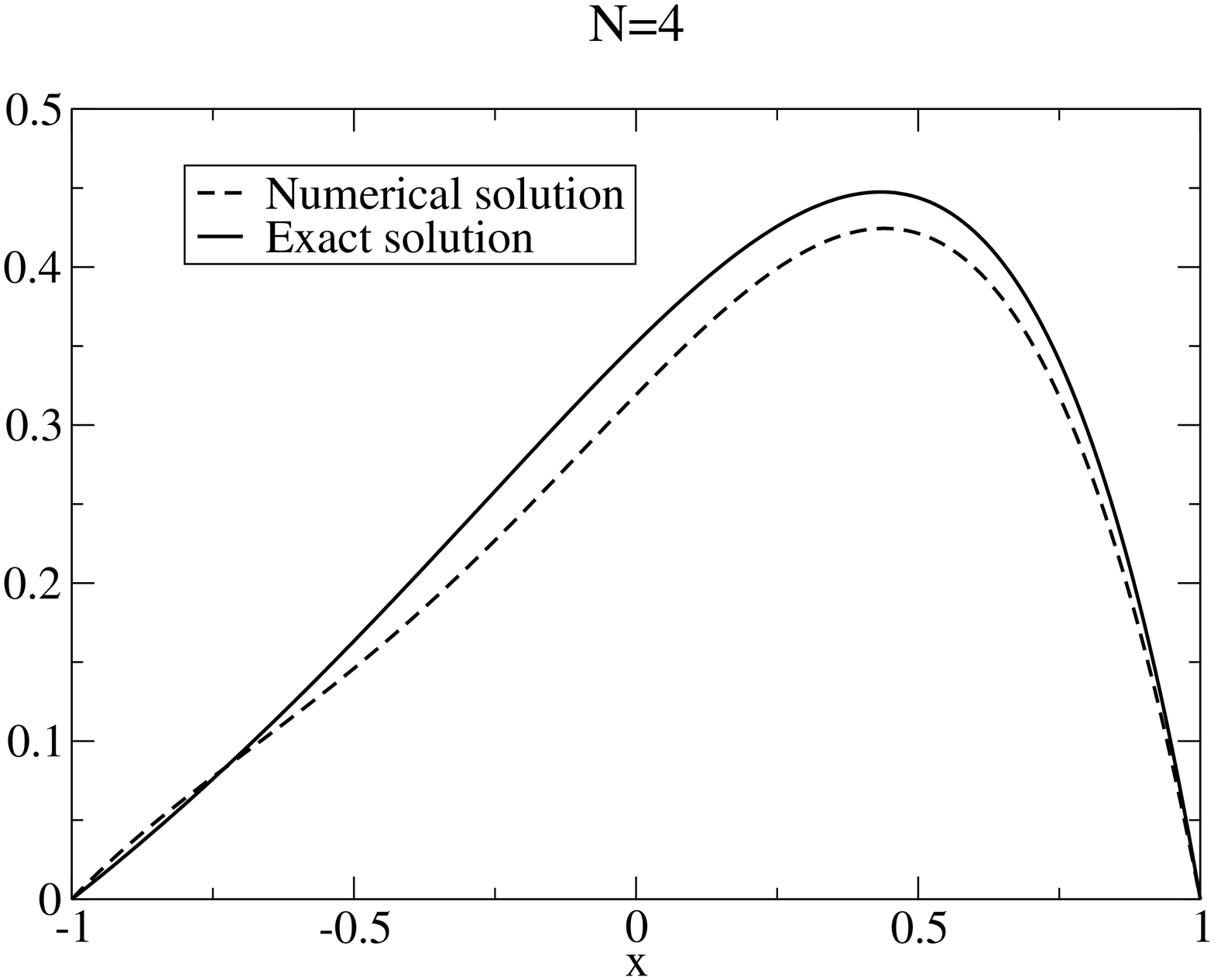}
\includegraphics[height=5.5cm]{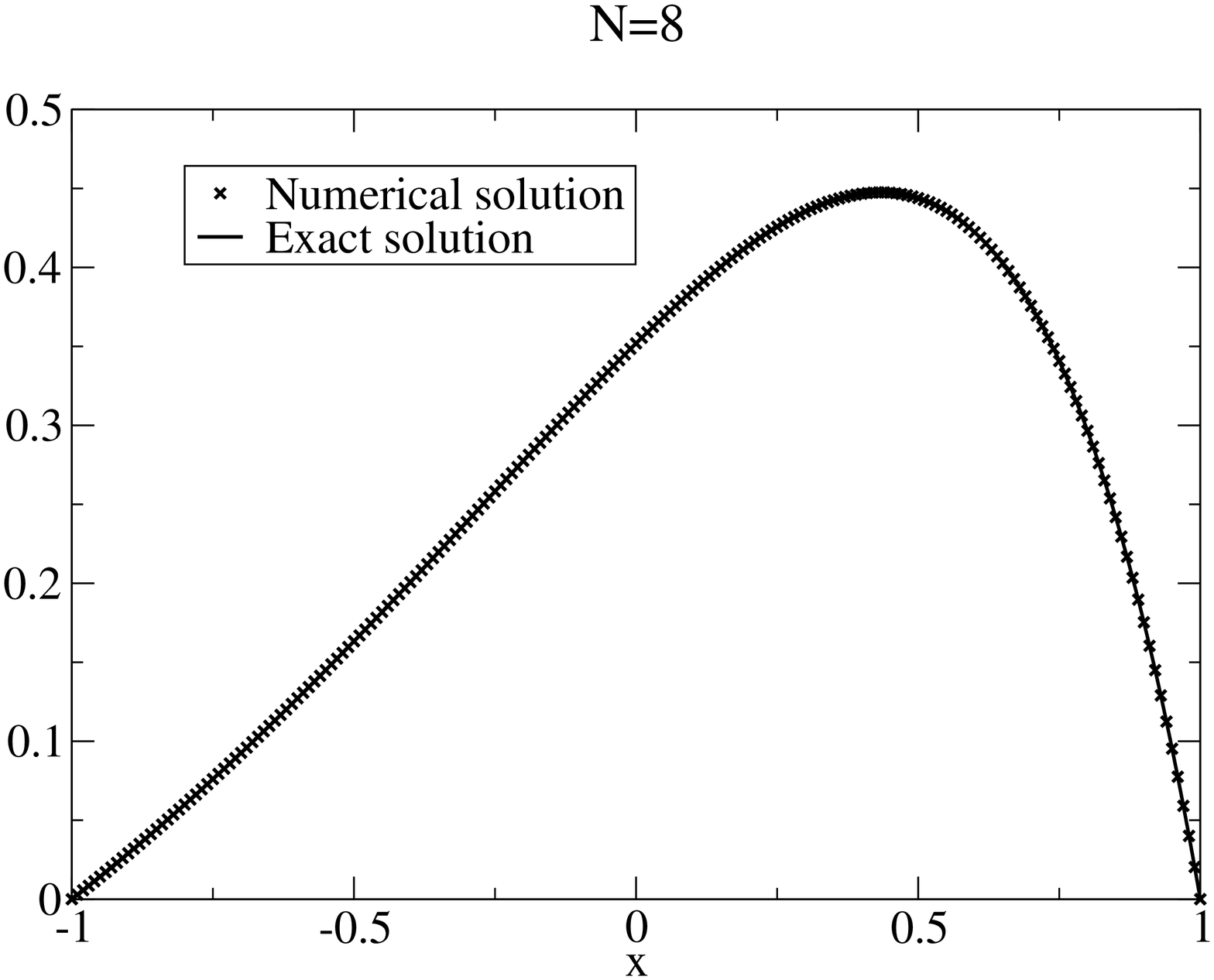}}
\caption{\label{f:galer}
Exact solution (solid line) and numerical one (dashed line and cross symbols), for the Galerkin method, for $N=4$ and $N=8$.
}
\end{figure}

In the case of our test problem, for $N=4$ the system is : 
\be
\l( \begin{tabular}{ccc}
$2 \pi$ &  $-4 \pi$&  $-4\pi$ \\
$8\pi$ &   $-8\pi$ & 0 \\
0  &  $8\pi$  &  $-26\pi$ \\
\end{tabular}\r) 
\l( \begin{tabular}{c}
$\tilde{u}^G_0$ \\
$\tilde{u}^G_1$ \\
$\tilde{u}^G_2$ \\
\end{tabular}\r) = 
\l( \begin{tabular}{c}
0.521 \\
-1.705 \\
0.103\\
\end{tabular}\r).
\ee
The Galerkin coefficients are $\tilde{u}^G_0 \simeq -0.160 ; \tilde{u}^G_1 \simeq -0.092 ; \tilde{u}^G_2 \simeq -0.029$ and the standard ones $\tilde{u_0}\simeq 0.189 ; \tilde{u_1}\simeq 0.092 ; \tilde{u_2} \simeq -0.160 ; \tilde{u_3}\simeq -0.092 ;
 \tilde{u_4}\simeq -0.029$. The numerical solution is compared to the exact one on Fig. 
 \ref{f:galer}.
 
\subsection{The methods are optimal}

\begin{figure}
\centerline{
\includegraphics[height=8cm]{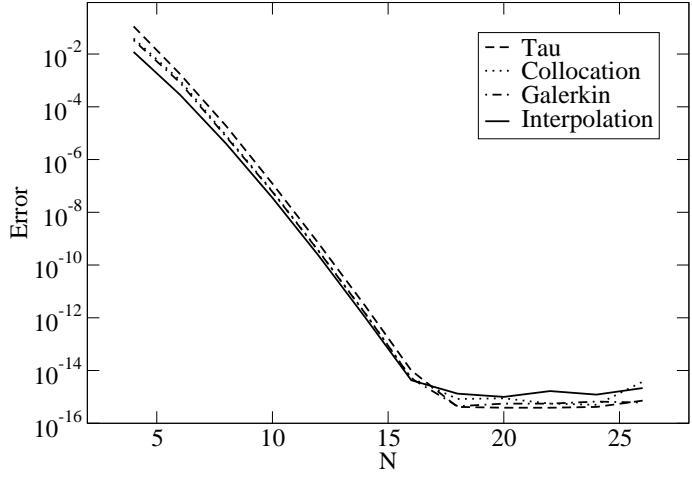}}
\caption{\label{f:errors_1d}
Maximum relative difference between the exact solution and the numerical one as a function of the number of coefficients, for the Tau-method (dashed line), the collocation one (dotted one) and the Galerkin method (dashed-dotted line). Is also shown, the relative difference between the exact solution and its interpolant.
}
\end{figure}

On Fig. \ref{f:errors_1d}, the maximum error between the numerical solution and the analytical one is shown, as a function of the number of coefficients $N$. The three methods exhibit similar features : an exponential decay before reaching the machine accuracy and being limited by round-off errors. All the functions involved in our test being ${\mathcal C}^\infty$, it is not surprising that the error is evanescent. Also plotted on Fig. \ref{f:errors_1d} is the maximum difference between the exact solution and its interpolant. The behavior is very similar to the three other curves. Thus, it seems that all the three resolution methods do not introduce more error than the one that would be already done when interpolating the exact solution.

This notion can be made more precise and is called {\em optimality}. Let us assume that the exact solution of an equation, $u_{\rm exact}$, is known. One can compared it to its interpolant $I_N u_{\rm exact}$ and to the obtained numerical solution $u_{\rm num}$. The resolution method is then said to be optimal if and only if ${\rm max}\l(|u_{\rm exact}-I_N u_{\rm exact}|\r)$ and
${\rm max}\l(|u_{\rm exact}-u_{\rm num.}|\r)$ have the same behavior when $N \rightarrow \infty$. Of course, this is usually a difficult quality to assess, given that the exact solution is rarely known. However, on our particular simple example, all the three methods presented are optimal.

\section{Multi-domain solvers}
\subsection{Motivation}
As previously seen, spectral methods loose some of their appeal when they have to deal with non ${\mathcal C}^\infty$ functions. Indeed, due to the Gibbs phenomenon, the convergence of the numerical solution to the real one is drastically slowed down.

Nevertheless, they are a great deal of physical problems where such discontinuities occur. We can for example mention the shock front in fluid dynamics or the surface of a strange star where a jump of matter density occurs. From a mathematical point of view, it is also sometime useful to use different variables and functions in various regions of space, thus creating numerical discontinuities.

For all those reason, one has to introduce several numerical domains to cover the physical space of interest. If one can ensure that the various discontinuities lie at the boundaries, then, in each domain, one will deal only with ${\mathcal C}^\infty$ functions, thus recovering exponential convergence.

\subsection{Test problem}
As in Sec. \ref{s:oned}, the various multi-domain solvers are presented on a simple test problem. The physical space is the interval $-1 \leq x \leq 1$. On this interval, one considers the following equation : 
\be
\label{e:eq_disc}
-\frac{{\rm d}^2u}{{\rm d}x^2} + 4 u = S
\ee
with the boundary conditions $u\l(-1\r) = 0$ and $u\l(1\r) = 0$. The source $S$ is a step function : $S\l(x<0\r) = 1$ and $S\l(x>0\r) = 0$. Under those conditions, the analytical solution is given by :
\bea
u\l(x<0\r) &=& \frac{1}{4} - \l(\frac{e^2}{4}+ B^- e^4\r) \exp\l(2x\r) + B^- \exp\l(-2x\r) \\
u\l(x>0\r) &=&  B^+ \l(\exp\l(-2x\r) - \frac{1}{e^4} \exp\l(2x\r)\r),
\eea
where the constants are 
\begin{itemize}
\item $B^- = -\fraction{1}{8\l(1+e^2\r)} - \fraction{e^2}{8\l(1+e^4\r)}$
\item $B^+ = \fraction{e^4}{8}\l(\fraction{e^2}{\l(1+e^4\r)} - \fraction{1}{\l(1+e^2\r)}\r)$.
\end{itemize}
As for the one-domain case, this solution is not polynomial.

\begin{figure}
\centerline{
\includegraphics[height=8cm]{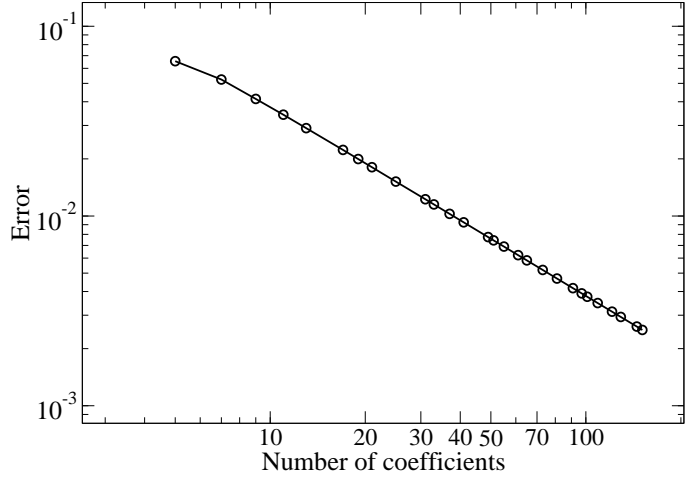}}
\caption{\label{f:tau_gibbs}
Error done when using a Tau-method with a discontinuous source, as a function of the number of coefficients. The convergence is no longer exponential but follows a power-law in $N^{-1}$.
}
\end{figure}

Of course, due to the discontinuity of the source, the solution is only ${\mathcal C}^1$ at $x=0$. This causes a great lose of accuracy for the one-domain solvers presented in Sec. \ref{s:oned}. For example, the error made when solving Eq. (\ref{e:eq_disc}) with the standard Tau-method is presented on Fig. \ref{f:tau_gibbs}. As expected, even if one converges to the right solution, the decay of the error is no longer exponential but only follows a power-law (like $N^{-1}$ in this particular case). For a practical point of view, the lose of accuracy is spectacular : with as many coefficients as $N=100$ one does not even reach a precision of $10^{-3}$. This is to be compared with the results of Sec.\ref{s:oned}, where machine precision could be reached with only 25 coefficients (see 
Fig. \ref{f:errors_1d}).

In order to cope with this problem, one can introduce two subdomains of the interval : 
\begin{itemize}
\item Domain 1 : $x\leq0$ described by the variable $x_1 = 2x+1$ ($-1\leq x_1\leq 1$).
\item Domain 2 : $x\geq 0$ described by $x_2 = 2x-1$ ($-1 \leq x_2 \leq 1$).
\end{itemize}
In each subdomain, the functions are expanded on the basis polynomials, with respect to the auxiliary variable. For instance, in domain 1, the solution is given by : 
\be
u\l(x<0\r) = \sum_{i=0}^N \tilde{u}^1_i T_i \l(x_1\l(x\r)\r).
\ee
When implementing various operators, one has to remember that one is working with the auxiliary variables. For example $\fraction{\rm d}{dx} = 2 \fraction{\rm d}{dx_i}$.

\subsection{A multi-domain Tau-method}
In each domain, one writes the usual residual equations for the Tau-method (see \ref{s:tau}). For instance, in the domain 1, this gives raise to 
\be
\l(T_n, R\r) = 0 \Longrightarrow \sum_{i=0}^N L_{ni} \tilde{u}^1_i = \tilde{s}^1_n.
\ee
This is a set of $N+1$ equations among which one has to relax the last two, in order to impose various boundary and continuity conditions.  $L$ is the differential operator expressed with respect to $x_1$. The same is done in the second domain.

As for the usual Tau method, the boundary conditions are enforced by two additional equations. In order to close the problem, one has to impose that the solution and its first derivative with respect to $x$ are continuous at the boundary between the two domains : 
\begin{itemize}
\item $u^1\l(x_1=1, x=0\r) = u^2\l(x_2=-1, x=0\r)$
\item $\fraction{{\rm d}u^1}{{\rm d}x}\l(x_1=1, x=0\r) = 
\fraction{{\rm d}u^2}{{\rm d}x}\l(x_2=-1, x=0\r)$
\end{itemize}

We are left with a system of : $2N-2$ residual equations, $2$ boundary conditions and $2$ matching conditions, for a total of $2N+2$ equations. The unknowns are the coefficients of the solution in both domains which are in the right number $2N+2$. For our test problem, when using $N=4$ in both domains, 
the system to be inverted is : 
\be
\l( \hspace{-0.2cm} \begin{tabular}{cccccccccc}
4 &  0 &  -16 & 0 & -128 & 0 & 0 & 0 & 0 & 0 \\
0 & 4 & 0 & -96 & 0 & 0 & 0 & 0 & 0 & 0 \\
0  & 0 & 4 & 0  & -192 & 0 & 0 & 0 & 0 & 0 \\
1 & -1 & 1 & -1 & 1 & 0 & 0 & 0 & 0 & 0 \\
1 & 1 & 1 & 1 & 1 & -1 & 1 & -1 & 1 & -1 \\
0 & 1 & 4 & 9 & 16 & 0 & -1 & 4 & -9 & 16 \\
0 & 0 & 0 & 0 & 0 & 1 & 1 & 1 & 1 & 1 \\
0 & 0 & 0 & 0 & 0 & 4 & 0 & -16 & 0 & -128 \\
0 & 0 & 0 & 0 & 0 & 0 & 4 & 0 & -96 & 0 \\
0 & 0 & 0 & 0 & 0 & 0 & 0 & 4 & 0 & -192 \\
\end{tabular}\hspace{-0.2cm} \r) \hspace{-0.2cm} 
\l(\hspace{-0.2cm}  \begin{tabular}{c}
$\tilde{u}^1_0$ \\
$\tilde{u}^1_1$ \\
$\tilde{u}^1_2$ \\
$\tilde{u}^1_3$ \\
$\tilde{u}^1_4$ \\
$\tilde{u}^2_0$ \\
$\tilde{u}^2_1$ \\
$\tilde{u}^2_2$ \\
$\tilde{u}^2_3$ \\
$\tilde{u}^2_4$ \\
\end{tabular}\hspace{-0.2cm} \r)\hspace{-0.2cm}  = \hspace{-0.2cm} 
\l(\hspace{-0.2cm}\begin{tabular}{c}
1 \\
0 \\
0 \\
0 \\
0 \\
0 \\
0 \\
0 \\
0 \\
0 \\
\end{tabular} \hspace{-0.2cm} \r)
\ee
The solution is then given in both domains by 
\begin{itemize}
\item 
$\tilde{u}^1_0 \simeq 0.082 ; \tilde{u}^1_1 \simeq 0.044 ; \tilde{u}^1_2 \simeq -0.036 ; 
\tilde{u}^1_3 \simeq  0.0018 ; \tilde{u}^1_4 \simeq -0.0007$
\item  
$\tilde{u}^2_0 \simeq  0.038; \tilde{u}^2_1 \simeq  -0.044; \tilde{u}^2_2 \simeq 0.008 ; 
\tilde{u}^2_3 \simeq -0.0018 ; \tilde{u}^2_4 \simeq 0.00017$
\end{itemize}
The convergence properties of this method will be presented in Sec. \ref{s:compare_multi}.

\subsection{Homogeneous solution method}
This method is the closest to the analytical way of solving ordinary differential equations in various domains. The idea is just to compute some particular solution in both domains, to get the homogeneous solutions and to do the appropriate linear combination to fulfill the boundary conditions and continuity requirements.

Most of the time, the homogeneous solution are known analytically. Generically, their number is given by the order of the operator $L$ of the equation. In our particular example, $L$ is of second order and one gets $2$ homogeneous solutions in each domain. One has to compute the coefficients of those solutions with respect to the auxiliary variables $x_1$ and $x_2$. In our case, one finds : 
\begin{itemize}
\item $h_1\l(x\r) = \exp\l(2x\r) = \exp\l(x_1-1\r) = \exp\l(x_2+1\r)$
\item $h_2 \l(x\r) = \exp\l(-2x\r) = \exp\l(-x_1+1\r) = \exp\l(-x_2-1\r)$.
\end{itemize}
For instance, for $h_1$ this gives raise to the following coefficients, in both domains :
\begin{itemize}
\item in the domain 1 : $\l(0.47 ; 0.42 ; 0.1 ; 0.017 ; 0.0020 \r)$
\item in the domain 2 : $\l(3.4 ; 3.1 ; 0.74 ; 0.12 ; 0.015\r)$
\end{itemize}

In order to get one particular solution in each domain, one can use whatever method already presented in the one-domain case (see Sec. \ref{s:oned}). The conditions enforced at the boundaries of each domain have no importance because they will be modified by linear combination with the homogeneous solutions. For instance, using a Tau-method in each domain, and demanding that the first two coefficients of the particular solutions are zero (equivalent to imposing two boundary conditions) gives the following system in the first domain (which reduces to a $\l(N-1\r) \times \l(N-1\r)$ system) :
\be
\l( \begin{tabular}{ccccc}
1 & 0 & 0 & 0 & 0 \\
0 & 1 & 0 & 0 & 0\\
4 & 0 & -16 & 0 & -128 \\
0 & 4 & 0 & -96 & 0 \\
0 & 0 & 4 & 0 & -192 \\

\end{tabular}\r) 
\l( \begin{tabular}{c}
$\tilde{g}_0$ \\
$\tilde{g}_1$ \\
$\tilde{g}_2$ \\
$\tilde{g}_3$ \\
$\tilde{g}_4$ \\
\end{tabular}\r) = 
\l( \begin{tabular}{c}
0 \\
0 \\
1 \\
0 \\
0 \\
\end{tabular}\r)
\ee
which enables us to obtain the following particular solution, in domain 1 $\tilde{g}_0 = 0 ; \tilde{g}_1 = 0 ; \tilde{g}_2 \simeq -0.53 ; \tilde{g}_3 = 0 ; \tilde{g}_4 =  -0.001$.

So, in each domain, the solution is given by the sum of the particular solution $g$ and the two homogeneous ones $h_1$ and $h_2$, with unknowns coefficients $\alpha^i$ and $\beta^i$, 
\be
u\l(x\r) = g^i\l(x_i\r) + \alpha^i h^i_1 \l(x_i\r) + \beta^i h^i_2 \l(x_i\r)
\ee
where $i$ denotes the domain.

In order to determine the appropriate values of the $4$ coefficients $\alpha^i$ and $\beta^i$, one has to impose : 
\begin{itemize}
\item Boundary condition at $x=-1$ i.e. at $x_1=-1$ 
$$  g^1\l(-1\r) + \alpha^1 h^1_1 \l(-1\r) + \beta^1 h^1_2 \l(-1\r) = 0$$
\item Boundary condition at $x=1$ i.e. at $x_2= 1$
$$g^2\l(1\r) + \alpha^2 h^2_1 \l(1\r) + \beta^2 h^2_2 \l(1\r) = 0$$
\item Matching of the solution at $x=0$, i.e. at $x_1=1$ and $x_2=-1$
$$ g^1\l(1\r) + \alpha^1 h^1_1 \l(1\r) + \beta^1 h^1_2 \l(1\r) = g^2\l(-1\r) + \alpha^2 h^2_1 \l(-1\r) + \beta^2 h^2_2 \l(-1\r)$$
\item Matching of the first derivative at $x=0$, i.e. at $x_1=1$ and $x_2=-1$
$$g'^1\l(1\r) + \alpha^1 h'^1_1 \l(1\r) + \beta^1 h'^1_2 \l(1\r) = g'^2\l(-1\r) + \alpha^2 h'^2_1 \l(-1\r) + \beta^2 h'^2_2 \l(-1\r)$$.
\end{itemize}

If the problem is well-posed, this system of $4$ equations admits a unique solution for the $4$ coefficients of the homogeneous solutions, thus allowing us to construct the solution of the equation in all space. For instance, in our particular case, for $N=4$, the matching system is 
\be
\l( \begin{tabular}{cccc}
7.39 &  0.135 &  0 & 0 \\
1 & 1 & -1 & -1 \\
1  &  -1  & -1 & 1 \\
0 & 0 & 0.135 & 7.39 \\
\end{tabular}\r) 
\l( \begin{tabular}{c}
$\beta^1$ \\
$\alpha^1$ \\
$\beta^2$ \\
$\alpha^2$ \\
\end{tabular}\r) = 
\l( \begin{tabular}{c}
0.055 \\
0.055 \\
-0.23 \\
0 \\
\end{tabular}\r)
\ee
which admits the solution : 
$\beta^1 = 0.0048 ; \alpha^1 = 0.14; \beta^2 = 0.094 ; \alpha^2 = -0.0017 $. The coefficients of the numerical solution, in each domains are then :
\begin{itemize}
\item 
$\tilde{u}^1_0 \simeq 0.083 ; \tilde{u}^1_1 \simeq 0.044 ; \tilde{u}^1_2 \simeq -0.036 ; 
\tilde{u}^1_3 \simeq  0.0018 ; \tilde{u}^1_4 \simeq -0.0008$
\item  
$\tilde{u}^2_0 \simeq  0.038; \tilde{u}^2_1 \simeq  -0.044; \tilde{u}^2_2 \simeq 0.008 ; 
\tilde{u}^2_3 \simeq -0.0018 ; \tilde{u}^2_4 \simeq 0.00016$.
\end{itemize}
Numerical accuracy of this method will be shown in Sec. \ref{s:compare_multi}.

\subsection{Variational formulation}
Contrary to all the methods previously presented, the variational one is only easily applicable when one is using Legendre polynomials. Indeed, as will be seen later, it requires that the measure is 
$w\l(x\r)=1$.

The starting point is the usual residual equation, written here explicitly in its integral form
\be
\l(\xi, R\r) = 0 \Longrightarrow \int \xi \l(-u'' + 4 u\r) {\rm d} x = \int \xi S  {\rm d} x.
\ee
The term involving $u''$ is then integrated by parts, which gives : 
\be
\label{e:variation}
\l[-\xi u'\r] + \int \xi' u' {\rm d} x + 4 \int \xi u {\rm d} x = \int \xi S  {\rm d} x.
\ee
The method is easier to implement when, as for the collocation method, the test functions are : 
 $\xi_n =\delta\l(x-x_n\r)$, being zero but at one collocation point.
 
 In the following, it will also be convenient to define the derivative operator $D$ in the 
configuration space. It relates
the values of the function at each collocation point to the ones of its first derivative via :
\be
g'\l(x_k\r) = \sum_{j=0}^N D_{kj} g\l(x_j\r).
\ee
The matrix associated to $D$ is easily constructed. For example, for $N=4$ and Legendre polynomials, one gets : 
\be
D_{ij} = \l( \begin{tabular}{ccccc}
-10  &  13.5 &  -5.33 & 2.82 & -1  \\
-2.48  & 0 & 3.49 & -1.53 & 0.52 \\
0.75 &  -2.67  &  0 & 2.67 & -0.75 \\
-0.52  &  1.53  &  -3.49 & 0 & 2.48\\
1  &  -2.82  & 5.33 & -13.5 & 10  \\
\end{tabular}\r)
\ee
Using this operator and the Gauss quadrature rule (\ref{gauss}), one can compute the integrals involved in Eq. (\ref{e:variation}) :
\bea
\int \xi_n' u' {\rm d}x &=& \sum_{i=0}^N \xi_n' \l(x_i\r) u'\l(x_i\r) w_i = 
\sum_{i=0}^N \sum_{j=0}^N D_{ij} D_{in} w_i u\l(x_j\r) \\
\int \xi_n u {\rm d}x &=& \sum_{i=0}^N \xi_n\l(x_i\r) u\l(x_i\r) w_i = u\l(x_n\r) w_n\\
\int \xi_n s {\rm d}x &=& \sum_{i=0}^N \xi_n\l(x_i\r) S\l(x_i\r) w_i = S\l(x_n\r) w_n.
\eea
For the points {\em strictly inside} each domain, the integrated term $\l[-\xi u'\r]=0$ so that the 
residual equations take the following form : 
\be
\label{e:residu_var}
\sum_{i=0}^N \sum_{j=0}^N D_{ij} D_{in} w_i u\l(x_j\r) + 4 u\l(x_n\r) w_n = S\l(x_n\r) w_n, \quad 0<n<N
\ee
which is a set of $2N-2$ equations.

The situation of the point lying at the boundary between the two domains is a bit tricky. Indeed the same physical point can be described either : 
\begin{itemize}
\item as the last point of domain 1 : $n=N$ and $\l[-\xi u'\r] = - u'^{1} \l(x_1 = 1 ; x=0 \r)$ so that the residual equation is : 
\be
u'^{1} \l(x_1 = 1\r) = \sum_{i=0}^N \sum_{j=0}^N D_{ij} D_{iN} w_i u^1\l(x_j\r) + 
4 u^1\l(x_N\r) w_N - S^1\l(x_N\r) w_N.
\ee
\item as the first point of domain 2 : $n=0$ and $\l[-\xi u'\r] = u'^{2} \l(x_2 = -1 ; x=0 \r)$ and the residual equation is then 
\be
u'^{2} \l(x_2 = -1\r) = -\sum_{i=0}^N \sum_{j=0}^N D_{ij} D_{i0} w_i u^2\l(x_j\r) - 4 u^2\l(x_0\r) w_0 + S^2\l(x_0\r) w_0.
\ee
\end{itemize}
This duality can be used to impose implicitly the continuity of the first derivative of $u$ :
\bea 
\label{e:implicit_var}
\nonumber
&&u'^{1} \l(x_1 = 1 ; x=0 \r) = u'^{2} \l(x_2 = -1 ; x=0 \r) \Longrightarrow \\
\nonumber
&&\sum_{i=0}^N \sum_{j=0}^N D_{ij} D_{iN} w_i u^1\l(x_j\r) + 4 u^1\l(x_N\r) w_N + \sum_{i=0}^N \sum_{j=0}^N D_{ij} D_{i0} w_i u^2\l(x_j\r) + 4 u^2\l(x_0\r) w_0 \\
&&\phantom{xxxxxxxxxxxxxxxxxxxxxxxxxxxxxxx}= S^1\l(x_N\r) w_N + S^2\l(x_0\r) w_0.
\eea
The other equations are i) the boundary condition on the left $u^1\l(x_0\r) = 0$ ii) the boundary condition on the right $u^2\l(x_N\r) = 0$ and iii) the matching of the solution at the boundary between the two domains $u^1\l(x_N\r) = u^2\l(x_0\r)$. This is a total number of $2N+2$ equations for the $2N+2$ unknowns, which in this case, are the value of the solution at each collocation points, the $u^i\l(x_n\r)$.

For our example and $N=4$ in each domain, the system is : 
\be
\l( \hspace{-0.3cm}
\begin{tabular}{ccccccccc}
1 &  0 &  0& 0 & 0 & 0 & 0 & 0 & 0  \\
-16 & 28 & -12 & 2.5 & -0.7 & 0 & 0 & 0 & 0  \\
2.1  & -12 & 22 & -12  & 2.1 & 0 & 0 & 0 & 0  \\
-0.7 & 2.5 & -12 & 28 & -16 & 0 & 0 & 0 & 0  \\
0.2 & -0.7 & 2.1 & -16 & 29 & -16 & 2.1 & -0.7 & 0.2 \\
0 & 0 & 0 & 0 & -16 & 28 & -12 & 2.5 & -0.7  \\
0 & 0 & 0 & 0 & 2.1 & -12 & 22 & -12 & 2.1  \\
0 & 0 & 0 & 0 & -0.7 & 2.5 & -12 & 28 & -16  \\
0 & 0 & 0 & 0 & 0 & 0 & 0 & 0 & 1  \\
\end{tabular}\hspace{-0.3cm}\r)\hspace{-0.2cm}
\l(\hspace{-0.3cm}  \begin{tabular}{c}
$u^1\l(x_0\r)$\\
$u^1\l(x_1\r)$\\
$u^1\l(x_2\r)$\\
$u^1\l(x_3\r)$\\
$u^1\l(x_4\r) = u^2\l(x_0\r)$\\
$u^2\l(x_1\r)$\\
$u^2\l(x_2\r)$\\
$u^2\l(x_3\r)$\\
$u^2\l(x_4\r)$\\
\end{tabular}\hspace{-0.3cm} \r)\hspace{-0.2cm}  = \hspace{-0.2cm} 
\l(\hspace{-0.3cm}\begin{tabular}{c}
0 \\
0.54 \\
0.71 \\
0.54 \\
0.1 \\
0 \\
0 \\
0 \\
0 \\
\end{tabular} \hspace{-0.3cm} \r)
\ee
Once it is inverted, the solution is found in the configuration space :
\begin{itemize}
\item In domain 1 : $u^1 \l(x_n\r) = \l(0 ; 0.06 ; 0.12 ; 0.12 ; 0.092\r)$
\item In domain 2 : $u^2 \l(x_n\r) = \l(0.092 ; 0.064 ; 0.030 ; 0.0090 ; 0\r)$.
\end{itemize}
As for the other methods, the convergence properties are shown in the next section.

Before finishing with the variational method, it may be worthwhile to explain why Legendre polynomials were used. Suppose one wants to use Chebyshev polynomials. The measure is then 
$w\l(x\r) = \fraction{1}{\sqrt{1-x^2}}$. When one integrates the term in $u''$ by part one then gets 
\be
\int -u'' f w {\rm d}x = \l[-u' f w\r] + \int u' f' w' {\rm d} x
\ee
Because the measure is divergent at the boundaries, it is difficult, if not impossible, to isolate the term in $u'$. On the other hand this is precisely the term that is needed to impose the appropriate matching of the solution. There might be ways around this but this explains why the variational method has been presented with Legendre polynomials.

\subsection{Discussion}\label{s:compare_multi}

\begin{figure}
\centerline{
\includegraphics[height=8cm]{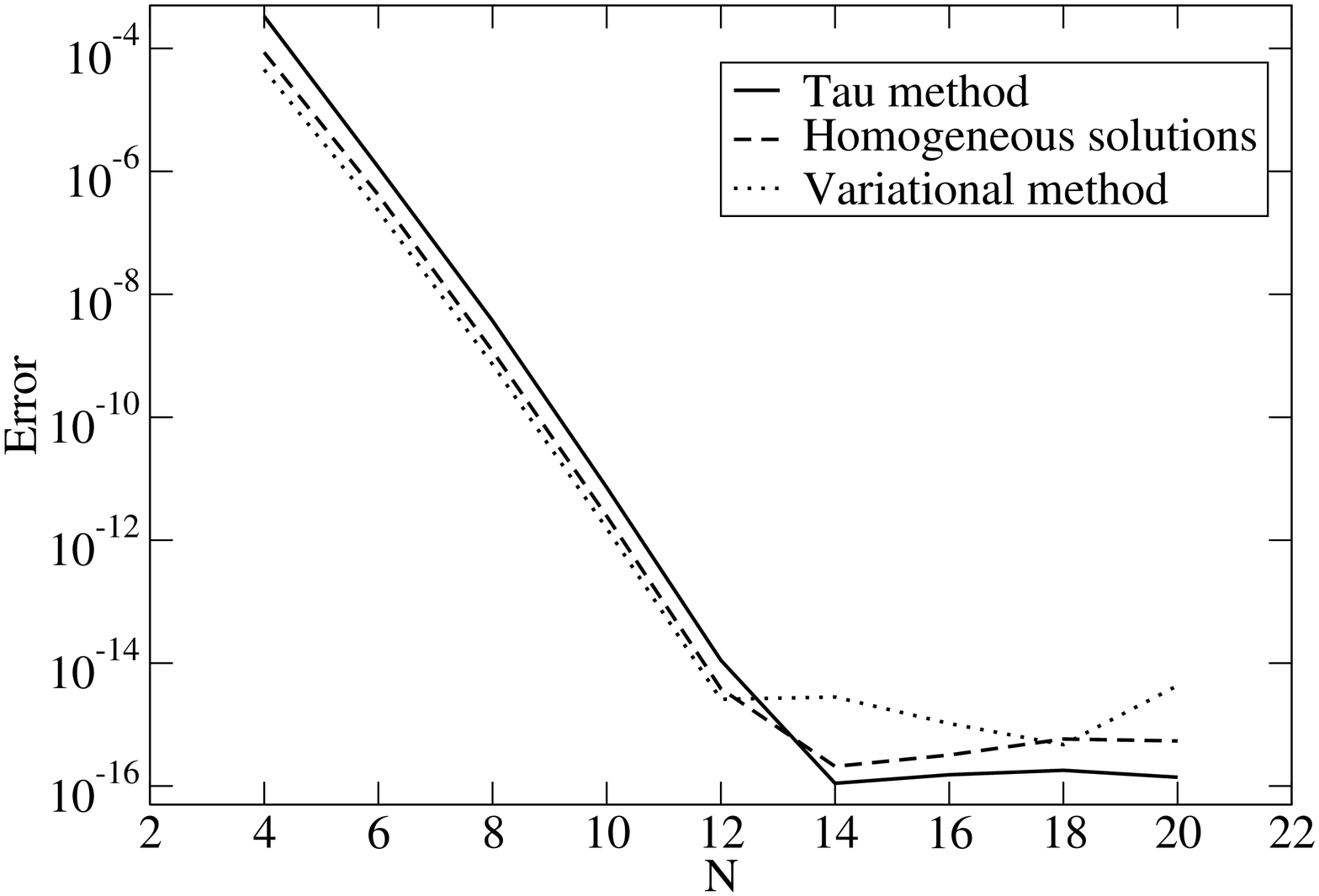}}
\caption{\label{f:errors_multi}
Relative difference between the analytical solution and the numerical one, for the three multi-domain methods exposed. The error is evanescent.
}
\end{figure}

On Fig. \ref{f:errors_multi}, we show the maximum relative difference between the numerical solution and the analytical one. Contrary to the case of just one domain (see Fig. \ref{f:tau_gibbs}), exponential decay of the error is recovered, as expected, and machine accuracy is very rapidly reached. This illustrates the usefulness of a multi-domain decomposition. 

From a numerical point of view, the method based on the homogeneous solutions is different from the two others. Indeed, when using it, the problem is treated domain by domain. The number of system to be solved is equal to the number of domains but each of those systems is roughly of size $N^2$ only. This is to be compared to the Tau-method and the variational one where only one global system is inverted but which is of size $\l(N N_{\rm D}\r)^2$. So, if the number of domain is important, one may consider the homogeneous solution method instead of the two others, which would give rise to huge systems.

On the other hand, the Tau-method and the variational one, do not require the knowledge of the homogeneous solutions which is a good point, especially when the differential operator $L$ is complicated. The variational method may seem a little more involved, especially as it requires to work with Legendre polynomials but, from a mathematical point of view, it is the only one that has been prove to be optimal.

\section{Conclusion}
Our introductory tour of spectral methods is now over. After having presented the mathematical foundations of this class of method, along with two families of standard polynomials, Legendre and Chebyshev ones, two simple test problems were treated by means of various methods. Three ways of solving an ordinary differential equation on a single domain were presented and tested : the Tau method, the collocation one and the Galerkin method. In the last part, a discontinuous source was treated by means of a multi-domain decomposition. Again, three methods were implemented and discussed : a multi-domain tau method, one based on the knowledge of homogeneous solutions and a variational method using Legendre polynomials. This simple examples were intended to illustrate the usefulness of spectral methods to reach very good accuracy with moderate computational resources.

However, this work is far from covering the huge field of all spectral methods. Among the many aspects that were ignored one can mention the use of Fourier transform for periodic functions. We also have left out all issues concerning compactification of space, by means of auxiliary variables like $1/r$. Three dimensional problems were also left out as were non-linear problems. Finally, we did not present any problem involving time evolution. Should the reader feel the urge to learn more about all this, we recommend that he consults some of the books given in the bibliography.


\end{document}